\documentclass[a4paper,fleqn,usenatbib]{mnras}
\pdfoutput=1

\usepackage[pdftex]{graphicx,xcolor}
\usepackage[T1]{fontenc}
\usepackage[utf8]{inputenc}
\usepackage{lmodern}
\usepackage{float}
\usepackage{caption}
\usepackage{tabularx}
\usepackage{amsmath, amssymb}
\usepackage{cuted}
\usepackage[english]{babel}
\usepackage{enumerate}
\usepackage[normalem]{ulem}
\usepackage{enumitem}
\usepackage{multirow}
\usepackage{booktabs}
\usepackage{textcase}
\usepackage{makecell}
\usepackage{xspace}

\usepackage[table]{xcolor}  
\usepackage{colortbl}       
\usepackage{array}          
\usepackage{orcidlink}
\usepackage{newtxtext,newtxmath}

\DeclareRobustCommand{\VAN}[3]{#2}
\let\VANthebibliography\thebibliography
\def\thebibliography{\DeclareRobustCommand{\VAN}[3]{##3}\VANthebibliography}

\def\be{\begin{equation}}
\def\ee{\end{equation}}

\newcommand{\dmigm}{{\rm DM}_{\rm IGM}}
\newcommand{\dmobs}{{\rm DM}_{\rm obs}}
\newcommand{\dmtot}{{\rm DM}_{\rm tot}}
\newcommand{\dmsmooth}{{\rm DM}_{\rm smooth}}
\newcommand{\dmhalo}{{\rm DM}_{\rm halo}}
\newcommand{\dmfil}{{\rm DM}_{\rm fil}}
\newcommand{\dmhost}{{\rm DM}_{\rm host}}
\newcommand{\dmmw}{{\rm DM}_{\rm MW}}
\newcommand{\avg}[1]{\left\langle #1 \right\rangle}
\newcommand{\turbofrb}{\texttt{turboFRB}\xspace}
\newcommand{\turboFRB}{\turbofrb}

\definecolor{valecol}{rgb}{0,0.5, 1.}

\title[Stochastic FRB DM model]{A stochastic forward model for the intergalactic dispersion-measure distribution of Fast Radio Bursts}

\author[Fortunato et al.]{
J\'eferson A. S. Fortunato\orcidlink{0000-0001-7983-1891},$^{1,2,3}$\thanks{jeferson.fortunato@uct.ac.za}
Valerio Marra\orcidlink{0000-0002-7773-1579},$^{4,5,6,8}$\thanks{valerio.marra@me.com}
\newauthor
Wiliam S. Hip\'olito-Ricaldi\orcidlink{0000-0002-1748-553X},$^{7,8}$\thanks{wiliam.ricaldi@ufes.br} and Amanda Weltman\orcidlink{0000-0002-5974-4114},$^{1,2,3}$\thanks{amanda.weltman@uct.ac.za}
\\
$^{1}$High Energy Physics, Cosmology \& Astrophysics Theory Group, Department of Mathematics and Applied Mathematics,\\ University of Cape Town, Cape Town 7700, South Africa\\
$^{2}$African Institute for Mathematical Sciences, 6 Melrose Road, Muizenberg, Cape Town 7945, South Africa\\
$^{3}$National Institute for Theoretical and Computational Sciences (NITheCS), Potchefstroom, South Africa\\
$^{4}$Departamento de Física, Universidade Federal de Ouro Preto, 35400-000, Ouro Preto, MG, Brazil\\
$^{5}$INAF -- Osservatorio Astronomico di Trieste, via Tiepolo 11, 34131 Trieste, Italy\\
$^{6}$IFPU -- Institute for Fundamental Physics of the Universe, via Beirut 2, 34151, Trieste, Italy\\
$^{7}$ Grupo de Física Teórica e Computacional, Departamento de Ci\^encias Naturais, CEUNES, Universidade Federal do Esp\'irito Santo, \\ Rodovia BR 101 Norte, km 60, CEP 29.932-540, S\~ao Mateus, ES, Brazil\\
$^{8}$N\'ucleo Cosmo-UFES, CCE, Universidade Federal do Esp\'irito Santo, Av. Fernando Ferrari, 540, CEP 29.075-910, Vit\'oria, ES, Brazil
}

\makeatletter
\def\@printed{Compiled using MNRAS \LaTeX\ style file v\@version}
\makeatother
\date{}

\begin{document}
\label{firstpage}
\pagerange{1--16}
\maketitle
\pubyear{2026}

\begin{abstract}
Fast Radio Bursts probe ionised baryons through their observed dispersion measures. We present \turbofrb, a semi-analytic stochastic forward model for the intergalactic dispersion-measure distribution, $P(\dmigm\mid z)$, that resolves the diffuse IGM, halo, and filament contributions as explicit physical channels, with the halo and filament encounter rates coupled by a latent line-of-sight environmental variable. Only four effective parameters are calibrated against hydrodynamical ray-traced IllustrisTNG benchmark. The model matches the benchmark mean DM to the percent level and yields a per-redshift Jensen--Shannon divergence of at most $5\times10^{-3}$ across $z = 0.5$--$2.5$. The per-sightline channel decomposition makes explicit what closed-form parametric descriptions cannot show: the diffuse IGM sets the body of the distribution, while halos and filaments populate the high-DM tail. Applied to representative localised FRBs, the forward likelihood quantifies host-excess events independently of their astrophysical signatures and recovers the injected $H_0$ within $1\sigma$ in a closed-loop consistency test. The \turbofrb package is available at \href{https://github.com/jefersonfortunato/turbofrb}{github.com/jefersonfortunato/turbofrb}.
\end{abstract}




\section{Introduction}
\label{sec:intro}

Fast Radio Bursts (FRBs) are millisecond radio transients that propagate across cosmological distances before reaching the observer~\citep{Lorimer2007,Petroff2019}. Because radio waves travel through ionised plasma with a frequency-dependent group velocity, every burst accumulates a dispersion delay proportional to the integrated free-electron column density along the line of sight. The corresponding observable is the dispersion measure (DM), and for a localised FRB it can be written as
\be
\dmobs = \dmmw + \dmigm + \frac{\dmhost}{1+z_{\rm host}},
\label{eq:dm_decomp}
\ee
where $\dmmw$ contains the Milky Way disk and halo free electron density contributions, $\dmigm$ is the intergalactic contribution, and $\dmhost$ is the source-host contribution evaluated in the host rest frame. In a homogeneous Universe one recovers the familiar Macquart relation $\avg{\dmigm(z)}\propto z$ at low redshift~\citep{Ioka2003,Inoue2004}, generalised in realistic cosmology by integrating the free-electron density along the light cone. This mean relation has already been used to infer baryon fractions, expansion histories, and tests of fundamental physics from localised FRBs~\citep{Macquart2020,James2022,Hagstotz2022,fortunato2025fast,fortunato2026}.

The mean relation is not sufficient for individual FRB sightlines. At fixed redshift, different lines of sight sample different environments: some pass mostly through diffuse intergalactic gas, while others intersect halo outskirts, groups, clusters, or cosmic filaments. The relevant observable is therefore the full conditional distribution
\be
P(\dmigm\mid z),
\label{eq:pdm_def}
\ee
which is broad, skewed, and non-Gaussian~\citep{bhandari2021probing}. Its body is set mainly by the diffuse IGM component, while halos and filaments populate the shoulder and high-DM tail. This distinction matters for likelihood analyses because Gaussian approximations or misestimated tails assign incorrect probabilities to high-DM events. As localised FRB samples grow in number, resolving the physical channels that shape $P(\dmigm\mid z)$ becomes increasingly important. This motivation goes back to ~\citet{McQuinn2014}, who showed that the shape of the DM distribution is sensitive to whether baryons reside inside halos or in more extended diffuse structures.

A widely used phenomenological description compresses the problem into a small number of effective parameters. In the Macquart-style picture, the normalised variable
\be
\Delta \equiv \frac{\dmigm}{\avg{\dmigm(z)}}
\label{eq:delta_def}
\ee
is described by a skewed one-parameter family with a redshift-dependent width usually written as
\be
\sigma_{\rm diff}(z) = \frac{F}{\sqrt{z}},
\label{eq:macquart_sigma}
\ee
with $F$ calibrated from localised FRBs~\citep{Macquart2020}. The corresponding fitting function for the PDF is written as \citep{McQuinn2014, ProchaskaZheng2019, james2022z}:
\be
P_{\rm diff}(\Delta)\propto \Delta^{-\beta}
\exp\!\left[
-\frac{\left(\Delta^{-\alpha}-C_0\right)^2}{2\alpha^2\sigma_{\rm diff}^2}
\right],
\label{eq:usual_pdf}
\ee
with $\alpha$, $\beta$, $C_0(z)$ and $\sigma_{\rm diff}(z)$ fitted to the IllustrisTNG simulation in \citet{Zhang2021}. This type of approximation is useful because it converts the full distribution into a small set of summary functions. It is also, however, known to have limitations. At low redshift the $F/\sqrt{z}$ scaling formally diverges, the fitted PDF family becomes less well behaved, and the identification of $\sigma_{\rm diff}$ with the true normalised scatter $\sigma_\Delta$ is no longer exact, as studied by \citet{Zhuge2026}. They also provided a complementary analytic viewpoint parametrized by halo-channel arguments,
\be
\sigma_\Delta^2(z)\sim
\frac{\sigma_{\rm halo}^2\,D_c(z)}{\avg{\dmigm(z)}^2\,l},
\label{eq:zhuge_sigma}
\ee
where $\sigma_{\rm halo}$ is the halo-induced scatter per encounter, $l$ is an effective spacing scale that absorbs cross-section and mass-function weighting, and $D_c$ is the comoving distance. This approach is physically suggestive, but it also leaves the key ingredients --- the scatter per encounter and the effective spacing --- to be supplied externally, and assumes implicitly that halo encounters dominate the variance budget at all redshifts.

Much of the recent FRB cosmology literature has relied on the TNG-calibrated analytic fit of ~\citet{Zhang2021} as a reference description of $P(\dmigm\mid z)$. ~\citet{Konietzka2025}, however, re-measured the TNG distribution using continuous mesh ray-tracing through Voronoi cells and showed that the original Zhang-based path integration suffers from a sparse-snapshot artefact that biases the width and higher moments of the distribution by tens of percent. In parallel, ~\citet{Torkamani2026} constructed an analytic baryonification model for $P(\dmigm\mid z)$, while ~\citet{ConnorFil2025} provided direct observational evidence that filament crossings can imprint measurable DM excesses. ~\citet{AndrewMasui} have additionally argued that, on linear scales, the FRB dispersion measure is an effectively unbiased tracer of the matter distribution, with bias parameter $b_{\rm DM}\simeq 1$. Taken together, these developments motivate a modelling framework that does not assume a closed-form PDF from the outset, but instead resolves the main physical channels that generate the distribution.

Here we present \turbofrb{}, a semi-analytic stochastic generator for $P(\dmigm\mid z)$ that builds each sightline as the sum of three physically motivated baryonic channels: a mean-preserving lognormal diffuse component, Poisson-sampled halo encounters integrated through NFW gas profiles \citep{navarro1997universal}, and Poisson-sampled filament encounters with finite-cylinder geometry, coupled by a latent environmental variable that correlates the structural channels. This explicit channel decomposition is the conceptual core of the model, distinguishing \turbofrb{} from both closed-form parametric descriptions and pure hydrodynamical ray-tracing: rather than assuming a PDF shape, the model resolves the physical mechanisms that generate it. The output is a per-sightline DM trace, from which PDFs, moments, quantiles, tail probabilities, and channel decompositions can be measured directly. Only four effective parameters are calibrated against the corrected TNG benchmark; the remaining inputs are fixed by external information, engine defaults, or budget closure. This makes \turboFRB a natural tool for Bayesian inference of cosmological parameters and for diagnosing individual FRB sightlines within a physically transparent framework.

The stochastic sampling architecture is inspired by the gravitational-lensing framework of ~\cite{kainulainen2009new,kainulainen2011accurate}, which demonstrated that line-of-sight PDFs can be built efficiently by Poisson-sampling halo encounters along cosmological sightlines and embedding them in large-scale structure. The FRB dispersion-measure problem shares this mathematical structure, since $\dmigm$ is likewise a line-of-sight integral of a positive-definite quantity that accumulates discrete contributions from overdense structures. The physical kernels, gas profiles, environmental modulation, and calibration strategy are, however, all specific to the dispersion-measure observable: weak-lensing convergence weights projected mass through a geometric lensing kernel,  whereas DM weights free-electron column density with a cosmological $(1+z)^{-1}$ dilution factor, requiring a complete reformulation of the encounter kernels and normalisation scheme.

Throughout, we assume a flat $\Lambda$CDM cosmology with $H_0=67.74$~km\,s$^{-1}$\,Mpc$^{-1}$, $\Omega_m=0.3089$, $\Omega_b=0.0486$, $\Omega_\Lambda=0.6911$, and the canonical Macquart partition $f_{\rm IGM,0}=0.84$. This is the cosmology adopted by the corrected ray-traced catalogue of ~\citet{Konietzka2025} which serves as our calibration target. Matching the simulation cosmology ensures that the calibrated effective parameters capture genuine differences in the conditional distribution shape rather than absorbing a residual cosmological-parameter offset between the model and the benchmark.

The paper is organised as follows. Section~\ref{sec:engine} defines the stochastic engine and calibration strategy. Section~\ref{sec:validation} validates the calibrated distributions against corrected TNG ray tracing. Section~\ref{sec:halostats} analyses the halo encounter statistics, Section~\ref{sec:outliers} applies the forward likelihood to representative FRBs, and Section~\ref{sec:h0} gives a closed-loop $H_0$ consistency test. Sections~\ref{sec:discussion} and~\ref{sec:conclusions} discuss the interpretation, limitations, and conclusions.

\section{The \turbofrb{} stochastic engine}
\label{sec:engine}

The forward model assembles the line-of-sight DM contribution as the sum of three physically motivated channels:
\be
\dmigm(z, \mathrm{LoS}) =
\dmsmooth(z) + \dmhalo(z, E) + \dmfil(z, E),
\label{eq:dm_channels}
\ee
where the smooth-IGM (diffuse), halo, and filament terms are defined below. The latent variable $E$ captures the line-of-sight environmental overdensity and couples the two structural channels through their encounter rates; the diffuse channel is drawn independently of $E$.

\subsection{Analytic mean dispersion measure}
\label{ssec:analytic_mean}

Before describing the stochastic channels, we specify the deterministic mean relation that all of them are anchored to. The analytic ensemble-mean intergalactic dispersion measure as a function of source redshift is
\be
\avg{\dmigm(z)} \;=\;
\int_{0}^{z}\!
\frac{c\,n_{b,0}\,f_e(z')\,(1+z')}{H(z')}\, dz',
\label{eq:dm_mean_analytic}
\ee
where $c$ is the speed of light, $H(z) = H_0\sqrt{\Omega_m(1+z)^3 + \Omega_\Lambda}$ is the Hubble parameter, and
\be
n_{b,0} \;=\; \frac{\rho_{b,0}}{m_p}
       \;=\; \frac{3 H_0^2\,\Omega_b}{8\pi G\,m_p}
\label{eq:nb0}
\ee
is the comoving baryon number density today. The dimensionless prefactor $f_e(z)$ is the fraction of cosmic baryons present as dispersing free electrons at redshift $z$, and is the quantity that encodes both the diffuse-IGM baryon partition and the time-dependent ionisation state of the plasma.

In the standard Macquart-style treatment~\citep{Macquart2020,deng2014cosmological}, $f_e$ is taken as approximately redshift-independent and factorised as $f_e^{\rm Macq} \simeq f_{\rm IGM,0}\,\chi_e$, with $f_{\rm IGM,0}=0.84$ the fraction of cosmic baryons in the diffuse IGM and $\chi_e = 1-Y_{\rm He}/2 \simeq 0.88$ the electron-per-baryon ratio of a fully ionised primordial plasma, giving $f_e \simeq 0.74$. ~\citet{Konietzka2025} have recently measured the corresponding ratio directly in IllustrisTNG, defining $f_{eb}(z)$ as the fraction of dispersing electrons to total baryons in the simulation volume. They show that $f_{eb}(z)$ is well approximated by a linear function,
\be
f_{eb}(z) \;\approx\; f_{eb,0} \;+\; f_{eb,1}\,z,
\qquad f_{eb,0}\approx 0.83,\;\; f_{eb,1}\approx 0.006,
\label{eq:fe_konietzka}
\ee
obtained from the TNG free-electron field; the mild positive evolution with redshift encapsulates the residual effects of stellar and AGN feedback together with the late stages of cosmic reionisation. 
Because ~\citet{Konietzka2025} provides the benchmark used throughout this study, all reported results use the linear fit~\eqref{eq:fe_konietzka} in Eq.~\eqref{eq:dm_mean_analytic}. A Macquart-style option with constant $f_e=\chi_e f_{\rm IGM,0} \approx 0.74$ is retained as an internally selectable alternative in the engine; this choice affects the mean relation but not the stochastic channel structure.

\subsection{Diffuse channel}
\label{ssec:diffuse}

The diffuse component $\dmsmooth$ represents the smooth ionised intergalactic medium between halos and filaments. It accounts for the variance in DM accumulated across the smooth low-density regions that constitute most of the comoving volume; while individual smooth-medium fluctuations are small, their integral over a $\sim$Gpc-scale path length contributes significantly to the broadening of $P(\dmigm\mid z)$.

We sample $\dmsmooth$ from a lognormal distribution with redshift-dependent log-scatter,
\be
\dmsmooth(z) \;=\; \mu_{\rm sm}(z) \cdot \exp\!\left[\sigma_{\rm sm}(z)\,X - \tfrac{1}{2}\sigma_{\rm sm}^2(z)\right],
\label{eq:smooth_lognormal}
\ee
where $X \sim \mathcal{N}(0,1)$ and 
\be
\sigma_{\rm sm}(z) \;=\; \sigma_0\,(1+z)^{\gamma_s}.
\label{eq:sigma_sm}
\ee
The mean-preservation factor $-\tfrac{1}{2}\sigma_{\rm sm}^2$ ensures $\avg{\dmsmooth(z)}=\mu_{\rm sm}(z)$ exactly, so the diffuse channel preserves its assigned budget share by construction and does not introduce an independent shift in the ensemble mean. The two free parameters of this channel are the amplitude $\sigma_0$ and the redshift exponent $\gamma_s$; both are treated as nuisance parameters and calibrated as described in Section~\ref{ssec:calibration}.

The lognormal form is adopted phenomenologically and is not the Macquart-style family of Eq.~\eqref{eq:usual_pdf}: it reflects the fact that the diffuse-medium DM is the integral of a positive-definite quantity (electron density times path length) with multiplicative fluctuations, while admitting an exact mean-preserving prescription. It is not a derivation: the underlying density field of the smooth IGM is closer to a Gaussian random field on large scales, and the integrated DM is approximately lognormal only in the limit where the integrand variance is dominated by the largest-scale modes. We retain the lognormal form because it is well-behaved at the low-DM tail and admits an exact mean-preserving prescription.

\subsection{Halo channel: Poisson encounters with NFW gas profiles}
\label{ssec:halos}

Halos along the LoS are drawn as a Poisson process with comoving differential rate\footnote{Throughout the paper we write redshift-evolution factors as $(1+z)^\eta$ for compactness. The engine internally implements these factors as $((1+z)/(1+z_{\rm piv}))^\eta$ with $z_{\rm piv}=1.0$. A different pivot rescales the amplitude parameters ($n_{\rm hbg}$, $\lambda_f^0$, $f_{h0}$, $f_{f0}$) and not the exponents; accordingly, all amplitudes quoted in this paper are defined at $z_{\rm piv}=1$. We keep the compact notation in the text and note this convention in the public code release.}
\be
\frac{d\Lambda_h}{dz}\;=\; n_{\rm hbg}\,(1+z)^{\eta_{n_h}}\,\avg{\sigma_{\rm cross}}_{M}\,\frac{c}{H(z)},
\label{eq:halo_rate}
\ee
where $n_{\rm hbg}$ is the effective calibrated comoving background halo number density, $\eta_{n_h}=-0.20$ controls its redshift evolution, and $\avg{\sigma_{\rm cross}}_M$ is the cross-section weighted average over the encounter mass function defined below; we use $\sigma_{\rm cross}(M)=\pi\,(x_{\rm max}\,R_{200}(M))^2$ with $x_{\rm max}=0.85$. The cumulative halo rate from the observer to source redshift $z_s$ is
\be
\Lambda_h(z_s)\;=\;\int_0^{z_s}\!\frac{d\Lambda_h}{dz}\,dz,
\label{eq:Lambda_h}
\ee
and is precomputed at construction time on the same redshift grid used for the analytic mean. Eq.~\eqref{eq:halo_rate} is an effective comoving parametrisation of the encounter rate: $R_{200}(M)$ is held fixed in redshift, so the cross-section carries no explicit $(1+z)$ scaling, and the proper-path and abundance-evolution factors that a fully physical rate would contain are absorbed into the calibrated $n_{\rm hbg}$ together with the fixed exponent $\eta_{n_h}$.

The individual halo masses along each LoS are drawn from a log-normal pool truncated to $M\in[10^{12},10^{14.8}]\,M_\odot$ with $\log_{10}M_{\rm mean}=12.65$ and $\log_{10}M_\sigma=0.35$; for each encounter, a single halo is selected from a pre-drawn pool of $25\,000$ candidates with probability proportional to its geometric cross-section $\propto R_{200}^2 \propto M^{2/3}$, which is the standard reweighting for unbiased Poisson sampling of a heterogeneous rate. The lower mass cut excludes sub-galactic halos that contribute negligibly to the DM, while the upper cut excludes the rarest cluster-mass halos which would otherwise be sampled too coarsely by the Poisson process. The virial radius scales with mass as $R_{200}(M) = R_{200}^{\rm ref}\,(M/M_{\rm ref})^{1/3}$ with $R_{200}^{\rm ref}=0.30$~Mpc at the pivot mass $M_{\rm ref}=10^{13}\,M_\odot$.

The encounter redshift of each halo along the LoS is drawn by inverse-CDF sampling against $\Lambda_h(z)$, ensuring that the distribution of encounter redshifts reproduces $d\Lambda_h/dz$ exactly without rejection.

For each halo encounter, the impact parameter $b$ is sampled uniformly within the cross-section disk, $b = x_{\rm max}\,R_{200}\sqrt{u}$ with $u\sim\mathcal{U}(0,1)$. The halo gas is assumed to trace the dark-matter density with a Navarro--Frenk--White~\citep{navarro1997universal} profile, regularised by a small core at the centre. Writing $u\equiv r/R_{200}$ and $u_s\equiv r_s/R_{200}=1/c_{200}$, the dimensionless gas shape function used by the engine is
\be
S(u)\;=\;\frac{1}{x(u)\,[1+x(u)]^2},
\qquad x(u)\;=\;\frac{u+\epsilon\,u_s}{u_s}\;=\;c_{200}\,u+\epsilon,
\label{eq:nfw_profile}
\ee
where $c_{200}=5.0$ is the adopted halo concentration \citep{Duffy2008} and $\epsilon=0.15$ is a small dimensionless core-regularisation parameter (in units of $r_s$) inserted to keep the central column finite. The volume integral of $S(u)$,
\be
J_{\rm mass}\;\equiv\;\int_0^1 4\pi\,u^2\,S(u)\,du,
\label{eq:Jmass}
\ee
sets the normalisation $\rho_0$ through the requirement that the total enclosed gas mass within $R_{200}$ equal $M_{\rm gas} = f_{\rm gas,h}\,M$, with an effective halo gas fraction $f_{\rm gas,h}=0.10$ that we treat as a fixed physical input. This value is consistent with the circumgalactic-medium constraints on $L_\star$ halos at $z\sim 0$ from ~\citet{ProchaskaZheng2019}, based on high-velocity clouds, O\,\textsc{vii} absorption, and the Large Magellanic Cloud sightline. To allow for halo-to-halo variation in gas content, the gas mass of each individual encounter is multiplied by a lognormal scatter factor with log-amplitude $0.12$, sampled independently per encounter.

The projected line-of-sight kernel at dimensionless impact parameter $x_b\equiv b/R_{200}$ is
\be
J_{\rm col}(x_b)\;\equiv\;2\!\int_0^{\sqrt{1-x_b^2}}\!
S\!\left(\sqrt{x_b^2+t^2}\right)\,dt,
\label{eq:nfw_projected_kernel}
\ee
and is precomputed on a grid of $x_b\in[0,x_{\rm max}]$ with 500 points, so a forward realisation only interpolates and does not re-evaluate the projection integral. The DM contribution of a single halo encounter at host redshift $z_{\rm enc}$, impact parameter $b$, and mass $M$ is then
\be
\dmhalo^{(1)}(M,b,z_{\rm enc})\;=\;\frac{\chi_e\,n_{b,0}^{\rm halo}(M)\,R_{200}(M)\,J_{\rm col}(x_b)}{1+z_{\rm enc}},
\label{eq:dm_halo_single}
\ee
where $\chi_e=0.88$ is the electron-per-baryon ratio adopted for fully ionised primordial gas, $n_{b,0}^{\rm halo}$ is the central baryon number density fixed by the gas-mass closure described above, and the factor $(1+z_{\rm enc})^{-1}$ is the cosmological dilution from the encounter rest frame to the observer.

The total halo channel along the line of sight is the sum over all Poisson-sampled encounters,
\be
\dmhalo(z_s,\mathrm{LoS})\;=\;\sum_{i=1}^{N_h}\dmhalo^{(1)}(M_i,b_i,z_i^{\rm enc}),
\label{eq:dm_halo_total}
\ee
where $N_h$ is drawn from a Poisson distribution whose effective rate includes the line-of-sight environmental modulation described in Section~\ref{ssec:environment}: $N_h\sim\mathrm{Poisson}\bigl[\Lambda_h(z_s)\,\xi_h(E)\bigr]$, with $\Lambda_h(z_s)$ from Eq.~\eqref{eq:Lambda_h} and $\xi_h(E)$ the halo environmental factor defined below. The parameter $n_{\rm hbg}$ absorbs the small offsets in mass-function shape, halo-gas content, and cross-section weighting between the engine's anchored prescription and the corrected TNG benchmark; it is the single degree of freedom in the halo channel that is calibrated rather than fixed by physical input. The remaining halo-channel parameters ($f_{h0}$, $\eta_h$, $\eta_{n_h}$, the cross-section weighting, the concentration $c_{200}$, and the effective gas fraction $f_{\rm gas,h}$) are all fixed by physical input at the values described above.

\subsection{Filament channel: finite-cylinder encounters}
\label{ssec:filaments}

Filament encounters are sampled with comoving differential rate
\be
\frac{d\Lambda_f}{dz}\;=\;\lambda_f^0\,(1+z)^{\eta_{\lambda_f}}\,\frac{c}{H(z)}\,\frac{1}{10^3\,{\rm Mpc/Gpc}},
\label{eq:fil_rate}
\ee
with $\lambda_f^0=0.45\,{\rm Gpc}^{-1}$ the comoving filament background rate at the pivot redshift, $\eta_{\lambda_f}=0.10$ its redshift exponent, and the factor of $10^{-3}$ converting from a per-Gpc rate to the per-Mpc convention used by the engine's path-length integration. The values of $\lambda_f^0$ and $\eta_{\lambda_f}$ are informed by the DESI/DisPerSE filament catalogue analyses of ~\cite{ConnorFil2025} and ~\cite{ConnorFilQuasar2024}. The cumulative filament rate $\Lambda_f(z_s)=\int_0^{z_s}d\Lambda_f/dz\,dz$ is precomputed on the same redshift grid as the halo rate.

Each filament is treated as a finite cylinder with radius and length that scale with its mass as
\be
R_f(M)=R_f^{\rm ref}\,(M/M_{\rm ref})^{0.20}\,,\qquad
L_f(M)=L_f^{\rm ref}\,(M/M_{\rm ref})^{0.15}\,,
\label{eq:fil_geom}
\ee
with $R_f^{\rm ref}=0.65\,$Mpc, $L_f^{\rm ref}=6.0\,$Mpc and $M_{\rm ref}=10^{13}\,M_\odot$, drawn from the empirical fits in the catalogue. The filament masses are drawn from a log-normal distribution truncated to $M\in[10^{11.8},10^{14.2}]\,M_\odot$ with $\log_{10}M_{\rm mean}=12.55$ and $\log_{10}M_\sigma=0.22$. The impact parameter $b$ is sampled uniformly within the cylinder cross-section disk ($b = R_f\sqrt{u}$ with $u\sim\mathcal{U}(0,1)$), and the cylinder orientation is sampled uniformly in $\cos\theta$, where $\theta$ is the inclination of the cylinder long-axis relative to the line of sight.

The radial baryon density inside the filament follows an isothermal $\beta$-model with $\beta=2/3$,
\be
n_b(b)\;=\;\frac{n_{b,0}^{\rm fil}}{1+\bigl(b/r_c\bigr)^2}\,,
\label{eq:beta_profile}
\ee
with core radius $r_c=0.18\,R_f$. The central density $n_{b,0}^{\rm fil}$ is fixed by requiring that the integral of the gas profile over the filament volume reproduce a gas mass $M_{\rm gas}=f_{\rm gas,f}\,M$, with an effective filament gas fraction $f_{\rm gas,f}=0.28$. The absolute normalisation of the filament channel is not set by this value directly but is fixed by the budget-enforcement step (Section~\ref{ssec:budget}); we adopt the isothermal $\beta$-model functional form ($\beta=2/3$) motivated by the DESI/DisPerSE filament analysis of ~\citet{ConnorFil2025}, while the core geometry ($r_c=0.18\,R_f$) is an engine choice. As with the halo channel, an independent lognormal scatter (log-amplitude $0.10$) is applied to the gas fraction of each individual filament encounter to capture filament-to-filament variation.

For an encounter at host redshift $z_{\rm enc}$ and impact parameter $b$, the projected DM contribution is then
\be
\dmfil^{(1)}(M,b,z_{\rm enc})\;=\;
\frac{\chi_e\,n_b(b)\,\ell(b)}{1+z_{\rm enc}}\,,
\label{eq:dm_fil_single}
\ee
where
\be
\ell(b)\;=\;\min\!\left[\,\frac{2\sqrt{R_f^2-b^2}}{\sin\theta},\;L_f\right]
\label{eq:chord}
\ee
is the chord length through the cylinder, clipped by the cylinder length when the inclination is small. Eq.~\eqref{eq:dm_fil_single} is an effective column kernel rather than an exact integration: the density is evaluated at the point of closest approach and multiplied by the chord length, instead of integrating $n_b[r(s)]$ along the oblique path, which would lower the column for inclined crossings. Because the budget step of Section~\ref{ssec:budget} fixes the ensemble mean of the channel, this approximation acts on the shape of the per-encounter distribution --- and hence mildly on the high-DM tail --- rather than on the channel normalisation. The total filament-channel contribution along the line of sight is the sum over all Poisson-sampled filament encounters, $\dmfil(z_s,\mathrm{LoS}) = \sum_{j=1}^{N_f}\dmfil^{(1)}(M_j,b_j,z_j^{\rm enc})$, with $N_f\sim\mathrm{Poisson}\bigl[\Lambda_f(z_s)\,\xi_f(E)\bigr]$, again with the environmental factor defined in Section~\ref{ssec:environment}.

\subsection{Environmental modulation}
\label{ssec:environment}

The latent environmental variable $E$ is drawn once per line of sight from a Gaussian distribution
\be
E\;\sim\;\mathcal{N}(0,\sigma_E^2),
\label{eq:E_draw}
\ee
and modulates the Poisson rates of the halo and filament channels through a mean-preserving exponential factor,
\be
\xi_X(E)\;=\;\exp\!\left[\,a_X\,E - \tfrac{1}{2}(a_X\sigma_E)^2\,\right],
\qquad X\in\{h,f\}.
\label{eq:env_factor}
\ee
The free parameter $\sigma_E$ controls the LoS-to-LoS amplitude of the modulation and is one of the four calibrated parameters. The two coupling exponents $a_h$ and $a_f$ control how strongly the halo and filament rates respond to the latent overdensity; we fix them at $a_h=0.60$ and $a_f=0.45$ as engine defaults, encoding the qualitative observation, drawn from cosmic-web simulations, that the halo abundance responds more sharply to the local overdensity than the filament population does.

The functional form~\eqref{eq:env_factor} is chosen so that the modulation is exactly mean-preserving,
\be
\avg{\xi_X(E)}_{E\sim\mathcal{N}(0,\sigma_E^2)} \;=\; 1,
\label{eq:env_mean_preserving}
\ee
which follows directly from the moment-generating function of a Gaussian. The $-\tfrac12(a_X\sigma_E)^2$ correction term therefore plays the same role for the environmental modulation as the analogous correction in the diffuse lognormal~\eqref{eq:smooth_lognormal}: it ensures that the modulation reshapes the LoS-to-LoS scatter without introducing a drift in the ensemble mean encounter rate.

Operationally, a single scalar $E$ is drawn per LoS and shared between the halo and filament channels. The effective Poisson rates that enter the per-LoS sampling are then
\be
\Lambda_h^{\rm eff}(z_s,E)=\Lambda_h(z_s)\,\xi_h(E),
\qquad
\Lambda_f^{\rm eff}(z_s,E)=\Lambda_f(z_s)\,\xi_f(E),
\label{eq:Lambda_eff}
\ee
and the encounter counts are drawn as
\be
N_h\sim\mathrm{Poisson}\bigl[\Lambda_h^{\rm eff}(z_s,E)\bigr],\qquad
N_f\sim\mathrm{Poisson}\bigl[\Lambda_f^{\rm eff}(z_s,E)\bigr].
\label{eq:Nh_Nf}
\ee
Because $\xi_X$ is mean-preserving in $E$, the ensemble-averaged rates over all sightlines satisfy $\avg{\Lambda_X^{\rm eff}} = \Lambda_X(z_s)$, so the budget rescaling described in Section~\ref{ssec:budget} is well-defined and the environmental channel adds only LoS-to-LoS variance to the marginal $P(\dmigm\mid z)$.

This is the mechanism by which the engine couples the halo and filament channels: along an overdense LoS ($E>0$), both $\xi_h(E)$ and $\xi_f(E)$ are simultaneously larger than unity, so more halos \emph{and} more filaments are intersected; along a void-like LoS ($E<0$), both factors are smaller than unity and the LoS samples a sparser population of structures. The asymmetry between $a_h$ and $a_f$ ensures that the halo rate responds more sharply to the latent overdensity than the filament rate does.

We acknowledge that the environmental modulation as implemented here is a simplification of the full cosmic-web structure: the latent $E$ is drawn independently per line of sight, with no explicit angular correlation, while in nature the environmental overdensity field has substantial coherence on Mpc scales.

\subsection{Budget enforcement}
\label{ssec:budget}

Given the channel weights $f_{h0}$ and $f_{f0}$, the smooth-channel weight is fixed by budget closure, $f_{s0}=1-f_{h0}-f_{f0}$. The channel weights evolve mildly with redshift through prescribed factors of the form $f_X(z)=f_{X0}\,(1+z)^{\eta_X}$, with the same compact-notation convention used elsewhere in the paper (see footnote in Section~\ref{ssec:halos}). The structural weights are additionally clipped to physical ranges ($f_h\in[0.03,0.22]$, $f_f\in[0.01,0.18]$, $f_h+f_f\le 0.90$) to prevent any unphysical excursion that could arise from an extreme combination of $\eta$ exponents; in practice, these clips are inactive at the calibrated configuration and across the calibration window. The smooth-channel mean is then
\be
\mu_{\rm sm}(z)\;=\;f_s(z)\,\avg{\dmigm(z)},
\ee
while the halo and filament channels are sampled stochastically according to their Poisson rates and gas profiles, and then \emph{rescaled} by a redshift-dependent multiplicative factor so that their ensemble means match the target budget shares,
\be
\begin{aligned}
\avg{\dmhalo(z)}=f_h(z)\,\avg{\dmigm(z)},\\
\avg{\dmfil(z)}=f_f(z)\,\avg{\dmigm(z)}.
\end{aligned}
\label{eq:budget_enforcement}
\ee
Operationally, the engine precomputes the raw ensemble mean of each stochastic channel,
\be
\avg{\dmhalo}^{\rm raw}(z_s)\;=\;\int_0^{z_s}\frac{d\Lambda_h}{dz}\,\avg{\dmhalo^{(1)}}_{M,b}(z)\,dz,
\label{eq:raw_mean_halo}
\ee
and similarly for the filaments, with the encounter-averaged DM contributions $\avg{\dmhalo^{(1)}}_{M,b}(z)$ estimated by Monte Carlo at five redshift anchors and linearly interpolated. The scale factors then read
\be
\begin{aligned}
s_h(z_s)\;=\;\frac{f_h(z_s)\,\avg{\dmigm(z_s)}}{\avg{\dmhalo}^{\rm raw}(z_s)},
\\
s_f(z_s)\;=\;\frac{f_f(z_s)\,\avg{\dmigm(z_s)}}{\avg{\dmfil}^{\rm raw}(z_s)},
\end{aligned}
\label{eq:scale_factors}
\ee
and the realised per-LoS halo and filament contributions are multiplied by $s_h$ and $s_f$ respectively before being summed. This step normalises the channels without distorting their conditional shape: the relative weights of common and rare encounters, of low-mass and high-mass halos, and of central and off-centre filament crossings are unchanged by the rescaling.

This distinction is important conceptually. The forward realisation for an individual LoS is still the stochastic sum of the smooth, halo, and filament contributions. What is constrained by the budget is the ensemble-average contribution of each channel at fixed redshift, rather than the addition of a fourth corrective term to every realisation. The generator is therefore anchored to the analytic mean relation through the rescaling of the smooth, halo, and filament components, rather than through a post-hoc shift to each LoS.

\begin{figure*}
\centering
\includegraphics[width=0.95\textwidth]{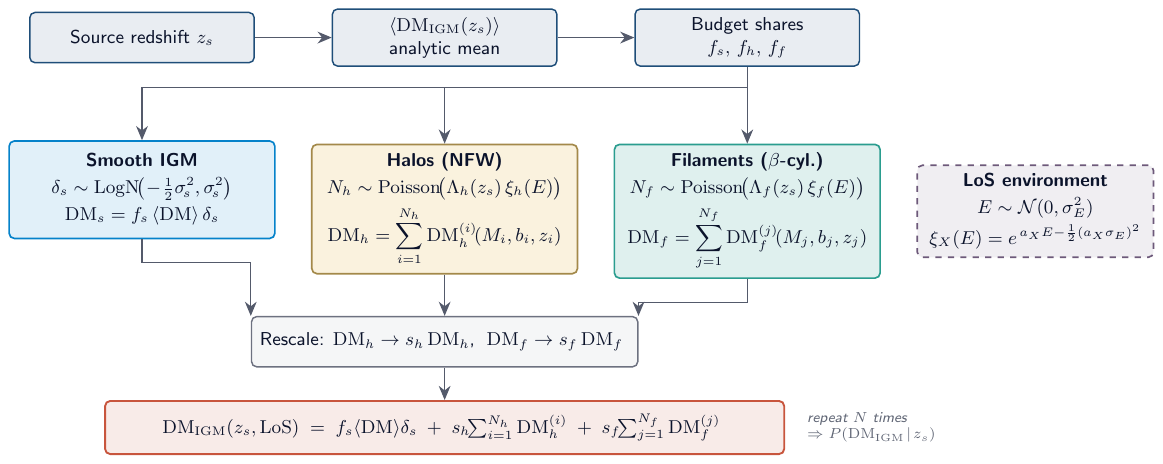}
\caption{Schematic of the \turbofrb forward sampler. For a source redshift $z_s$, the analytic mean $\langle\dmigm(z_s)\rangle$ is split into three channel budgets $(f_s, f_h, f_f)$. The smooth channel (\S\ref{ssec:diffuse}) draws a mean-preserving lognormal multiplier; the halo channel (\S\ref{ssec:halos}) Poisson-samples $N_h$ encounters at effective rate $\Lambda_h(z_s)\,\xi_h(E)$ and integrates NFW gas profiles along each chord; the filament channel (\S\ref{ssec:filaments}) Poisson-samples $N_f$ encounters at effective rate $\Lambda_f(z_s)\,\xi_f(E)$ and integrates $\beta$-model gas profiles through finite-cylinder geometry. A latent LoS environmental variable $E$ (\S\ref{ssec:environment}) couples the two structural channels through their Poisson rates. After sampling, the halo and filament sums are rescaled by $s_h, s_f$ (\S\ref{ssec:budget}) so that their ensemble means match the assigned budget shares without altering the conditional shape of each channel. The three contributions are summed to give one realisation of $\dmigm(z_s,\mathrm{LoS})$; repeating the procedure yields $P(\dmigm \mid z_s)$.} \label{fig:engine}
\end{figure*}

The full forward sampler is summarised in Fig.~\ref{fig:engine}. For each source redshift $z_s$, the engine reads the analytic mean $\avg{\dmigm(z_s)}$, partitions it into the three channel budgets, draws the latent environmental state $E$, samples each channel stochastically with its physical prescription, applies the budget rescaling described in this section, and returns the sum. Repeating the procedure yields the conditional distribution
$P(\dmigm \mid z_s)$.

\subsection{Parameter classification and calibration}
\label{ssec:calibration}

The sampling configuration exposes 15 scalar parameters, which are the quantities a user varies at run time; four of these are calibrated and the remaining hyperparameters --- mass pools and their dispersions, gas fractions and their scatters, $x_{\rm max}$, $c_{200}$, core and cylinder geometry --- are held fixed and catalogued in Appendix~\ref{app:fixed_params}. Following the principle that a forward generator should keep the calibrated subset to the minimum required to reproduce the target distribution, we classify them into three categories, summarised in Table~\ref{tab:params}.

\begin{table*}
\centering
\caption{Classification of the 15 scalar parameters exposed by the sampling configuration. \textbf{P}: fixed at the literature value or engine default. \textbf{D}: derived through budget closure. \textbf{N}: effective parameter calibrated by a single-shot grid search against the corrected ray-traced TNG benchmark of ~\citet{Konietzka2025} in $\Delta$-space, at matched (TNG) cosmology. The calibrated values reported in the rightmost column are the frozen four-parameter solution used throughout this paper.}
\label{tab:params}
\begin{tabular}{lccl}
\toprule
Parameter & Class & Value & Description / source \\
\midrule
$\sigma_0$           & N & $0.375$  & Diffuse log-scatter amplitude \\
$\gamma_s$           & N & $-1.460$ & Diffuse scatter redshift exponent \\
$\sigma_E$           & N & $0.450$  & Environmental log-scatter \\
$n_{\rm hbg}$        & N & $2.60\times 10^{-3}\,{\rm Mpc}^{-3}$ & Effective halo background density \\
\midrule
$f_{h0}$             & P & $0.10$  & Halo-channel weight at the pivot $z=1$ \\
$f_{f0}$             & P & $0.05$  & Filament-channel weight at the pivot $z=1$ \\
$\lambda_f^0$        & P & $0.45\,{\rm Gpc}^{-1}$  & Filament background rate from Mo~et~al. \\
$\eta_h$             & P & $+0.10$ & Halo-rate redshift exponent (engine default) \\
$\eta_f$             & P & $-0.10$ & Filament budget-weight redshift exponent (engine default) \\
$\eta_{n_h}$         & P & $-0.20$ & Halo-density redshift exponent \\
$\eta_{\lambda_f}$   & P & $+0.10$ & Filament-rate redshift exponent \\
$a_h$                & P & $0.60$  & Halo--environment coupling exponent \\
$a_f$                & P & $0.45$  & Filament--environment coupling exponent \\
spine concentration  & P & $1.0$   & Filament spine concentration factor \\
\midrule
$f_{s0}$             & D & $1-f_{h0}-f_{f0}$ & Smooth-channel weight by budget closure \\
\bottomrule
\end{tabular}
\end{table*}

The calibrated subset is therefore restricted to the four parameters $\{\sigma_0,\gamma_s,\sigma_E,n_{\rm hbg}\}$. The fixed parameters encode the adopted physical priors of the engine, while the derived weight $f_{s0}$ closes the budget once $f_{h0}$ and $f_{f0}$ are chosen. This classification limits the freedom of the calibration and keeps the model tied to its channel-level prescription. \turbofrb{} cannot freely fit an arbitrary benchmark: the stochastic channels remain tied to fixed physical inputs and engine defaults, and only four effective nuisance parameters are tuned. A model with enough free parameters could mimic almost any target distribution, whereas a model with four effective parameters and a fixed set of anchored physical inputs remains constrained by the encounter physics of the cosmic web. The provenance and physical role of the fixed engine inputs and modelling choices are summarised in Appendix~\ref{app:fixed_params}.

Before describing the calibration procedure, we explain the choice of comparison space. The two natural variables are the absolute DM and the normalised $\Delta = \dmigm/\avg{\dmigm}$. Calibrating directly in DM-space conflates two distinct sources of disagreement: a global normalisation difference in the mean DM budget and a genuine mismatch in the distribution shape. By calibrating in $\Delta$-space, where each distribution is normalised by its own sample mean, we isolate the second effect from the first. This is the conservative choice: it prevents the calibrator from absorbing a convention-level or budget-level offset that has nothing to do with the accuracy of the distribution shape itself. It also complements our decision to match the benchmark cosmology: with the cosmology fixed to the TNG values, the residual mean-DM offset is already at the percent level, so the $\Delta$-space normalisation removes only a very small residual and the calibration is genuinely shape-driven.

The four-parameter calibration is obtained with a two-stage grid search:
\begin{itemize}
\item \textbf{Target}: the ~\cite{Konietzka2025} continuous ray-traced TNG catalogue, comprising $1.2\times 10^5$ sightlines at each redshift bin, in the TNG cosmology.
\item \textbf{Comparison space}: $\Delta$-space, with each distribution normalised by its own sample mean.
\item \textbf{Metric}: the sum over $z\in\{0.5,1.0,1.5,2.0,2.5\}$ of the Jensen--Shannon divergence~\citep{lin1991divergence} computed from smoothed log-$\Delta$ kernel densities, with a bandwidth of $0.18$ in log space.
\item \textbf{Search}: a coarse grid scan over $\sigma_0\in[0.15,0.40]$, $\gamma_s\in[-1.30,-0.50]$, $\sigma_E\in[0.20,0.45]$, and $n_{\rm hbg}\in[1.0\times10^{-3},5.0\times10^{-3}]\,{\rm Mpc}^{-3}$, followed by a local refinement of five points per axis spanning $\pm$one coarse step around the coarse optimum. Because the coarse optimum for $\gamma_s$ falls on the lower edge of the coarse box, this refinement extends the search by one coarse step (of width $0.16$) below $-1.30$, which is how the reported value $\gamma_s=-1.460$ is reached.
\item \textbf{Reported solution}: $\sigma_0=0.375$, $\gamma_s=-1.460$, $\sigma_E=0.450$, $n_{\rm hbg}=2.60\times 10^{-3}\,\mathrm{Mpc}^{-3}$, with a per-redshift $\Delta$-space JSD at or below $\simeq 5\times 10^{-3}$ across the five-redshift calibration window (see Table~\ref{tab:konietzka_validation}).
\end{itemize}

Two of the four calibrated parameters --- $\gamma_s\simeq -1.46$ and $\sigma_E\simeq 0.45$ --- lie at the boundary of the searched region. Physically this reflects a steep redshift evolution of the diffuse log-scatter together with a sizeable LoS-to-LoS environmental amplitude, both needed to reproduce the simultaneous narrowing of the body at high redshift and the persistent extended tail. The two are not on the same footing, however. For $\sigma_E$ the refinement does probe beyond the coarse upper edge, and $\sigma_E=0.50$ returns a worse objective, so the adopted value is a genuine local optimum. For $\gamma_s$ the refinement returns the lowest value it explores ($\gamma_s=-1.46$, against $-1.38$ one step above), and smaller values were not scanned; we therefore do not claim a converged minimum in $\gamma_s$. Since $\gamma_s$ is a nuisance parameter of the diffuse channel entering only through the calibrated distribution shape, we adopt the four values as a frozen effective configuration rather than as a converged optimisation, and quote them without uncertainties throughout.

We therefore freeze the values quoted above as the calibrated configuration for the remainder of the paper.

\section{Validation against ray-traced TNG}
\label{sec:validation}

\begin{figure*}
\centering
\includegraphics[width=0.65\textwidth]{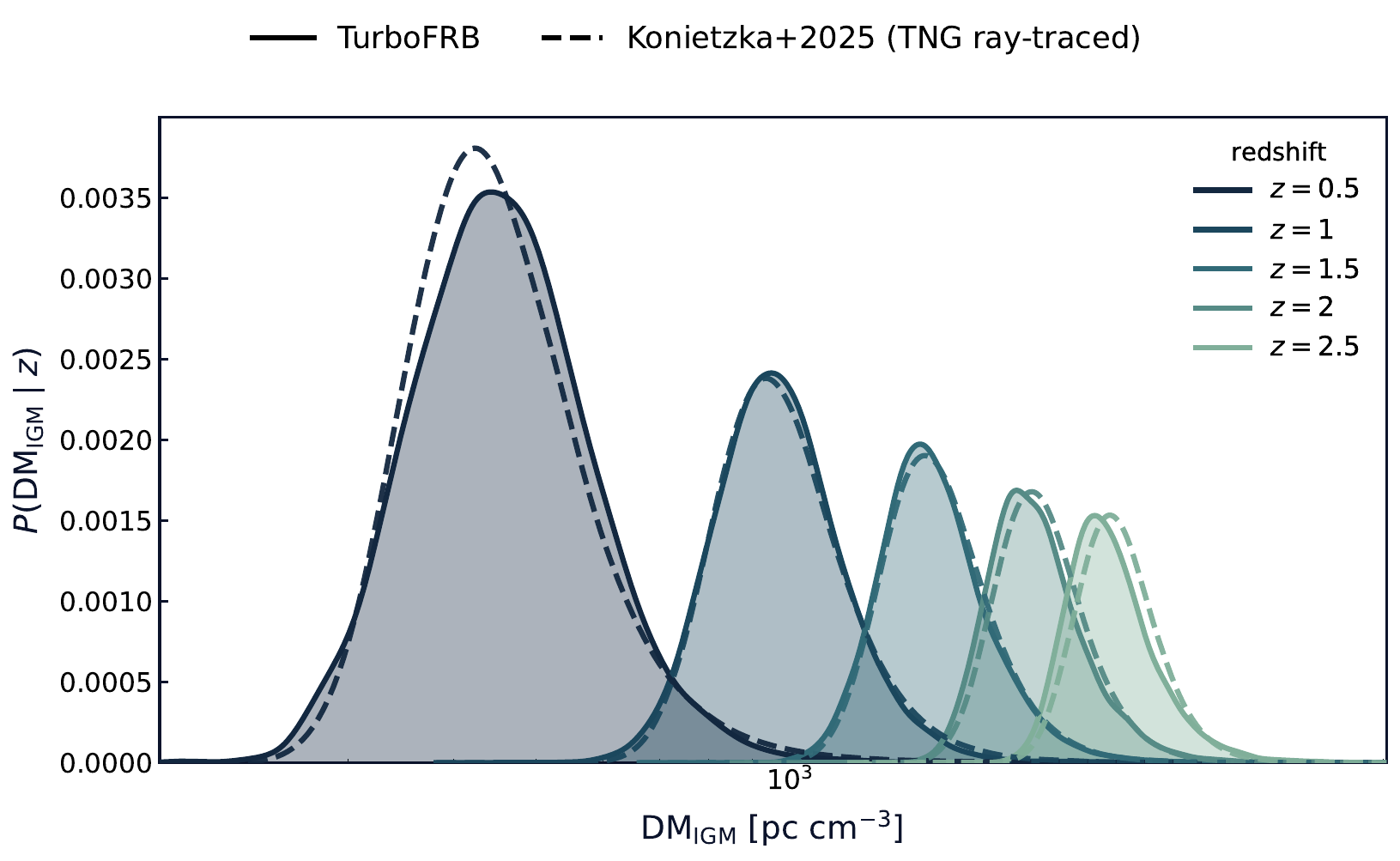}
\caption{Distribution of intergalactic dispersion measures $P(\dmigm\mid z)$ in DM-space at $z=0.5,1.0,1.5,2.0,2.5$. Solid filled curves show \turbofrb{} forward samples from the frozen 4-parameter calibration; dashed curves show the corrected ray-traced TNG measurements of \citet{Konietzka2025}. The broad skewed morphology is reproduced across the full range shown, and the DM-space mean is matched at the percent level. The remaining discrepancy is a modest excess high-DM tail of the forward model at the highest redshifts.}
\label{fig:fig1_dm}
\end{figure*}

The central prediction of the forward engine is the full conditional distribution $P(\dmigm\mid z)$ at fixed redshift. The first question, therefore, is not whether the model reproduces a single summary statistic, but whether it reproduces the shape of the corrected TNG distribution in the regime most relevant for localised FRBs. Figure~\ref{fig:fig1_dm} presents the DM-space comparison at five representative redshifts, $z\in\{0.5,1.0,1.5,2.0,2.5\}$. The \turbofrb{} ensemble at each redshift consists of $10^4$ forward realisations generated with the frozen 4-parameter calibration. The Konietzka comparison sample is the full corrected TNG catalogue at the corresponding redshift, comprising $1.2\times 10^5$ ray-traced sightlines from the public continuous catalogue. Both distributions are smoothed with a fixed log-DM kernel of bandwidth $0.18$, which is wide enough to suppress the per-bin noise of the estimator while preserving the shape features that distinguish the two prescriptions.

Several qualitative features are visible immediately. First, the two prescriptions agree on the broad morphology of the distribution: the peak is narrow and high at low redshift, the body broadens as the comoving distance grows, and a long positive-DM tail remains present at all redshifts. Second, the DM-space mean is aligned at the percent level. The mean ratio $\avg{\dmigm}^{\rm Turbo}/\avg{\dmigm}^{\rm Kon}$ stays in the range $0.982$--$0.990$ across $z=0.5$--$2.5$, so the residual differences are mainly shape differences rather than global shifts; \turbofrb{} sits marginally below Konietzka in mean DM at every redshift, by no more than $\simeq 2\%$. Third, the remaining discrepancy is most visible toward the lowest redshifts shown, where the \turbofrb{} distribution carries a slightly heavier high-DM tail than the benchmark.

\begin{table*}
\centering
\caption{Distribution-level comparison of \turbofrb{} against the ray-traced TNG measurements of \citet{Konietzka2025}, under the frozen four-parameter calibration. For each redshift we report the ensemble mean DM of both prescriptions, their ratio, and the two-sample KS and Jensen--Shannon statistics in both DM-space and $\Delta$-space (each distribution normalised by its own sample mean). Both prescriptions adopt the same TNG cosmology and the same $f_{eb}(z)$ electron fraction, so there is no convention-level offset between them: the $\Delta$-space metrics isolate shape disagreement from the residual percent-level difference between the analytic mean relation and the sample mean of each finite ensemble. These are the five redshifts used in the calibration, so the statistics are in-sample. The normalised scatter $\sigma_\Delta(z)$ is reported in Table~\ref{tab:halo_stats} and compared with external prescriptions in Fig.~\ref{fig:fig2_sigma_delta}.}
\label{tab:konietzka_validation}
\begin{tabular}{cccccccc}
\toprule
 & & & & \multicolumn{2}{c}{DM-space} & \multicolumn{2}{c}{$\Delta$-space} \\
\cmidrule(lr){5-6} \cmidrule(lr){7-8}
$z$ & $\langle\rm DM\rangle^{\rm Kon}$ & $\langle\rm DM\rangle^{\rm Turbo}$ & ratio & KS & JSD & KS & JSD \\
\midrule
0.50 & 505 & 500 & 0.990 & 0.042 & 0.0042 & 0.059 & 0.0047 \\
1.00 & 1035 & 1016 & 0.982 & 0.027 & 0.0021 & 0.038 & 0.0021 \\
1.50 & 1548 & 1531 & 0.989 & 0.031 & 0.0011 & 0.010 & 0.0005 \\
2.00 & 2032 & 2003 & 0.986 & 0.069 & 0.0033 & 0.023 & 0.0009 \\
2.50 & 2487 & 2451 & 0.986 & 0.098 & 0.0073 & 0.045 & 0.0030 \\
\bottomrule
\end{tabular}
\end{table*}

To quantify these visual impressions we compute, at each redshift, the two-sample Kolmogorov--Smirnov (KS) statistic on the raw samples and the Jensen--Shannon divergence \citep{lin1991divergence} on smoothed kernel-density estimates in both DM-space and $\Delta$-space. The KS statistic is sensitive to the largest cumulative distance between the empirical CDFs and is therefore a robust non-parametric measure of overall agreement; the JSD is sensitive to local density differences and to the body of the distribution where most of the probability mass resides. Reporting both gives a more complete picture than either alone. The values are summarised in Table~\ref{tab:konietzka_validation}: the $\Delta$-space JSD stays at or below $\simeq 5\times 10^{-3}$ at every redshift, with KS statistics of a few percent, and is smallest at $z\simeq 1.5$ where the two distributions are nearly indistinguishable within the smoothing kernel. These five redshifts are the ones used in the calibration, so Table~\ref{tab:konietzka_validation} quantifies in-sample goodness of fit rather than out-of-sample performance; the intermediate redshifts of Fig.~\ref{fig:fig2_sigma_delta} ($z=0.3,0.7,1.3,1.8,2.3$) were not used in the calibration and provide a complementary out-of-sample comparison of the normalised scatter. We also note that the percent-level agreement in mean DM is a mean-anchor check rather than an independent validation, since the analytic mean relation adopts the $f_{eb}(z)$ fit measured in the same simulation. We have verified that these distribution-level statistics are converged at the fiducial sample size of $10^4$ forward sightlines: increasing the ensemble to $5\times 10^4$ changes the measured $\sigma_\Delta(z)$ by less than the realisation-to-realisation scatter at fixed $N$. The convergence test is reported in Appendix~\ref{app:convergence}.

\begin{figure}
\centering
\includegraphics[width=1\columnwidth]{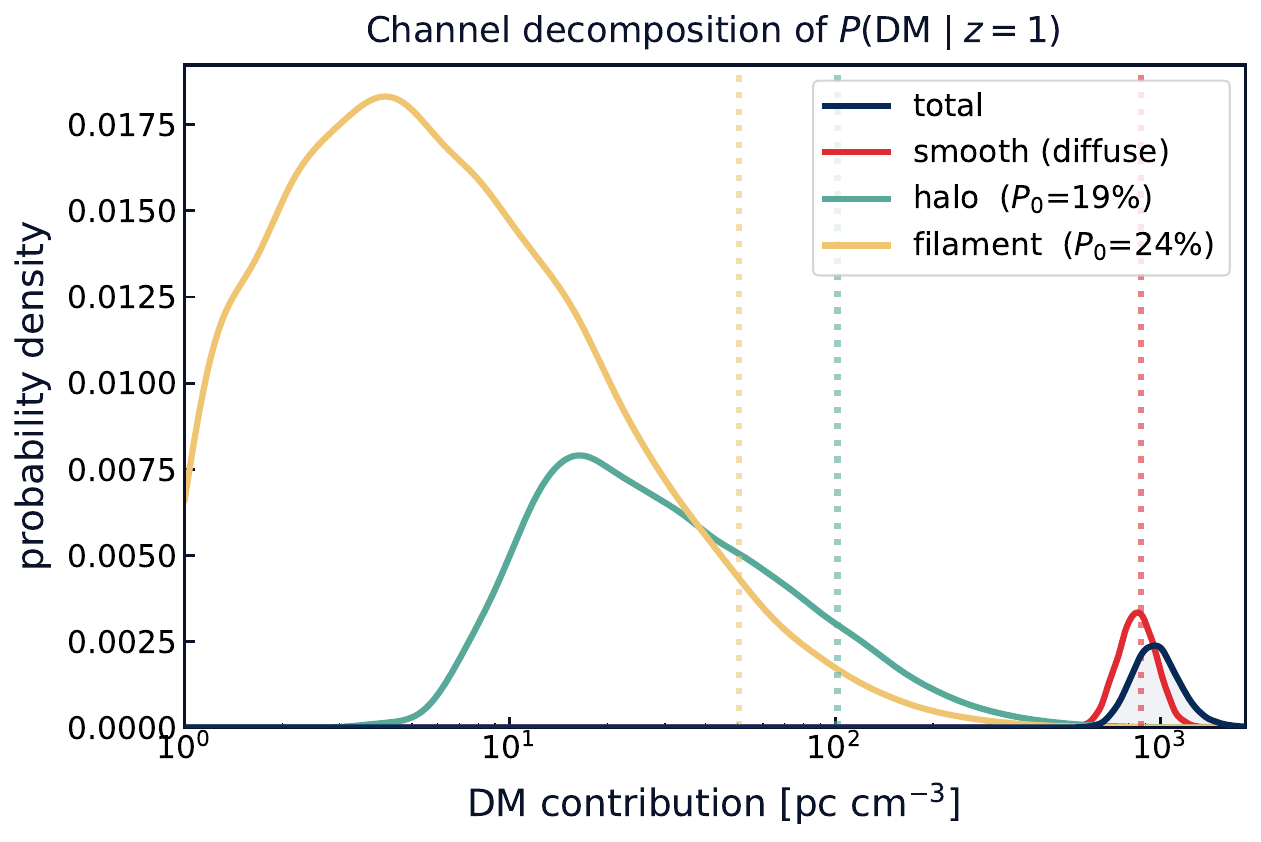}
\caption{Channel decomposition of $P(\dmigm\mid z{=}1)$ into its smooth (diffuse), halo, and filament contributions, measured directly from $2\times 10^4$ \turbofrb{} per-sightline traces under the calibration, shown on a logarithmic DM axis. A fraction of sightlines intersects no halo or filament, so these channels carry a finite probability of contributing exactly zero, which a logarithmic axis cannot display; the structural-channel densities are therefore conditional on ${\rm DM}>0$ and weighted by the corresponding probability, with the zero fractions $P_0$ quoted in the legend. The smooth channel sets the body of the distribution; the halo and filament channels are increasingly skewed and populate the shoulder and high-DM tail. Dotted vertical lines mark the channel means, computed over all sightlines including zero contributions.}
\label{fig:figB_decomp}
\end{figure}

Because the forward model is assembled from physically distinct channels, the same per-sightline traces that produce $P(\dmigm\mid z)$ can be decomposed into their smooth, halo, and filament contributions. Figure~\ref{fig:figB_decomp} shows this decomposition at $z=1$. The smooth diffuse channel dominates the body of the distribution, with a mean of $\simeq 872$~pc\,cm$^{-3}$ and a relatively symmetric core; the halo channel is strongly skewed, contributing a mean of $\simeq 102$~pc\,cm$^{-3}$ but a $95$th percentile of $\simeq 361$~pc\,cm$^{-3}$; and the filament channel is the smallest and most sharply skewed, with a mean of $\simeq 51$~pc\,cm$^{-3}$ and a median of only $\simeq 19$~pc\,cm$^{-3}$. The decomposition makes explicit what the single combined PDF cannot show directly: the diffuse channel sets the location and width of the body, while the halo and filament channels populate the shoulder and the extended high-DM tail. Within the adopted channel-budget prescription, the forward traces therefore show how each stochastic kernel populates the body and the tail of the total distribution. The shapes and the per-sightline realisations are outputs of the model; the mean partition between the three channels is prescribed by $f_{h0}$ and $f_{f0}$ through the budget step, so a good fit to the marginal $P(\dmigm\mid z)$ does not by itself single out this internal decomposition. Different weight choices could yield similar total PDFs with different channel splits.

\begin{figure}
\centering
\includegraphics[width=1\columnwidth]{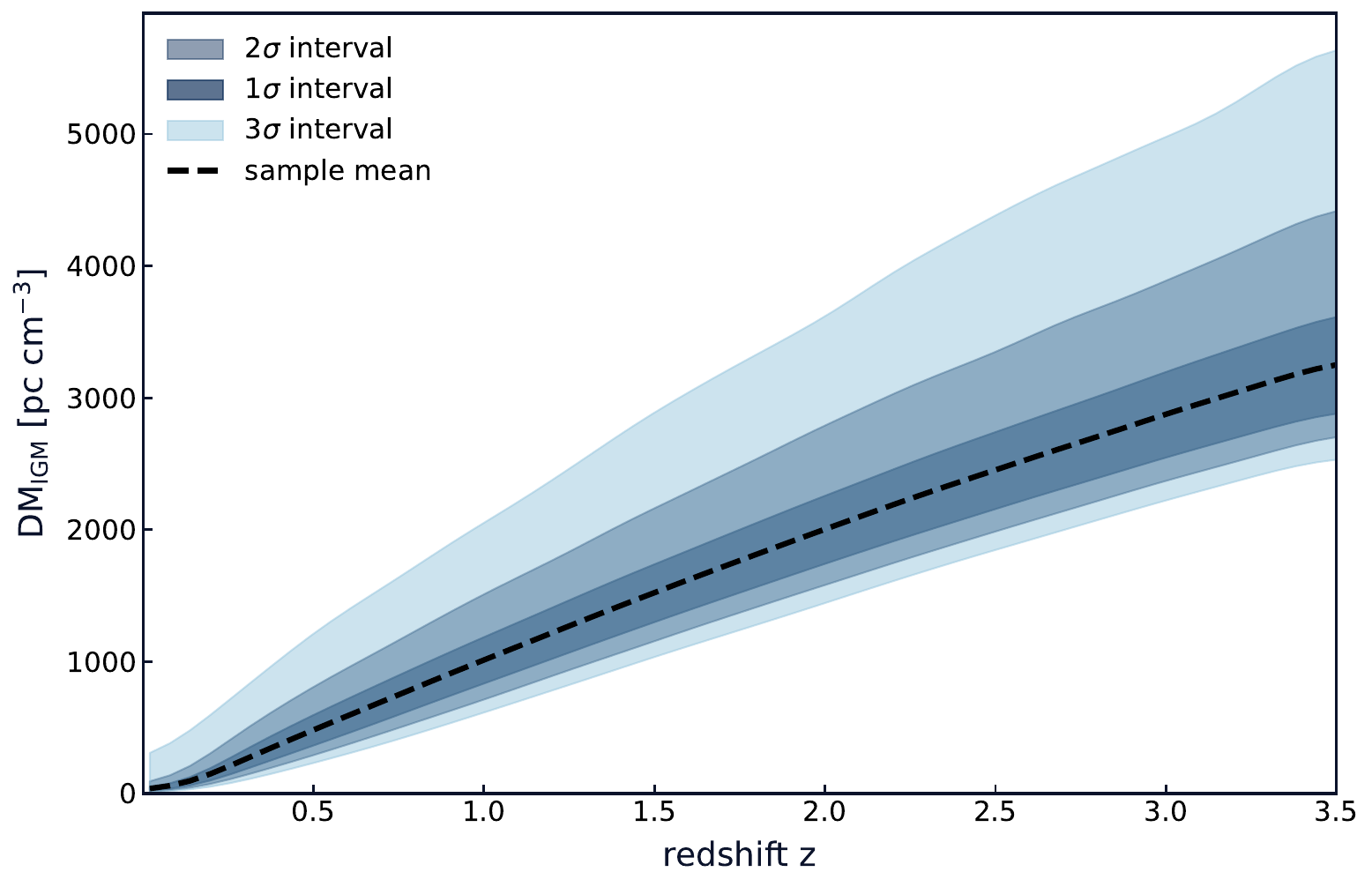}
\caption{Illustrative DM--$z$ diagram of the calibrated \turbofrb{} model, shown as percentile bands of $\dmigm(z)$ against host redshift. The figure visualises the width and skewness of the forward distribution across redshift.}
\label{fig:fig_dmz_fan}
\end{figure}

An illustrative DM--$z$ diagram showing the dispersion-measure quantiles of the calibrated model as continuous bands in the $(z,\dmigm)$ plane is shown in Figure~\ref{fig:fig_dmz_fan}. 

\subsection{Empirical scatter $\sigma_\Delta(z)$ and multi-prescription comparison}
\label{ssec:sigma_compare}

\begin{figure}
\centering
\includegraphics[width=1\columnwidth]{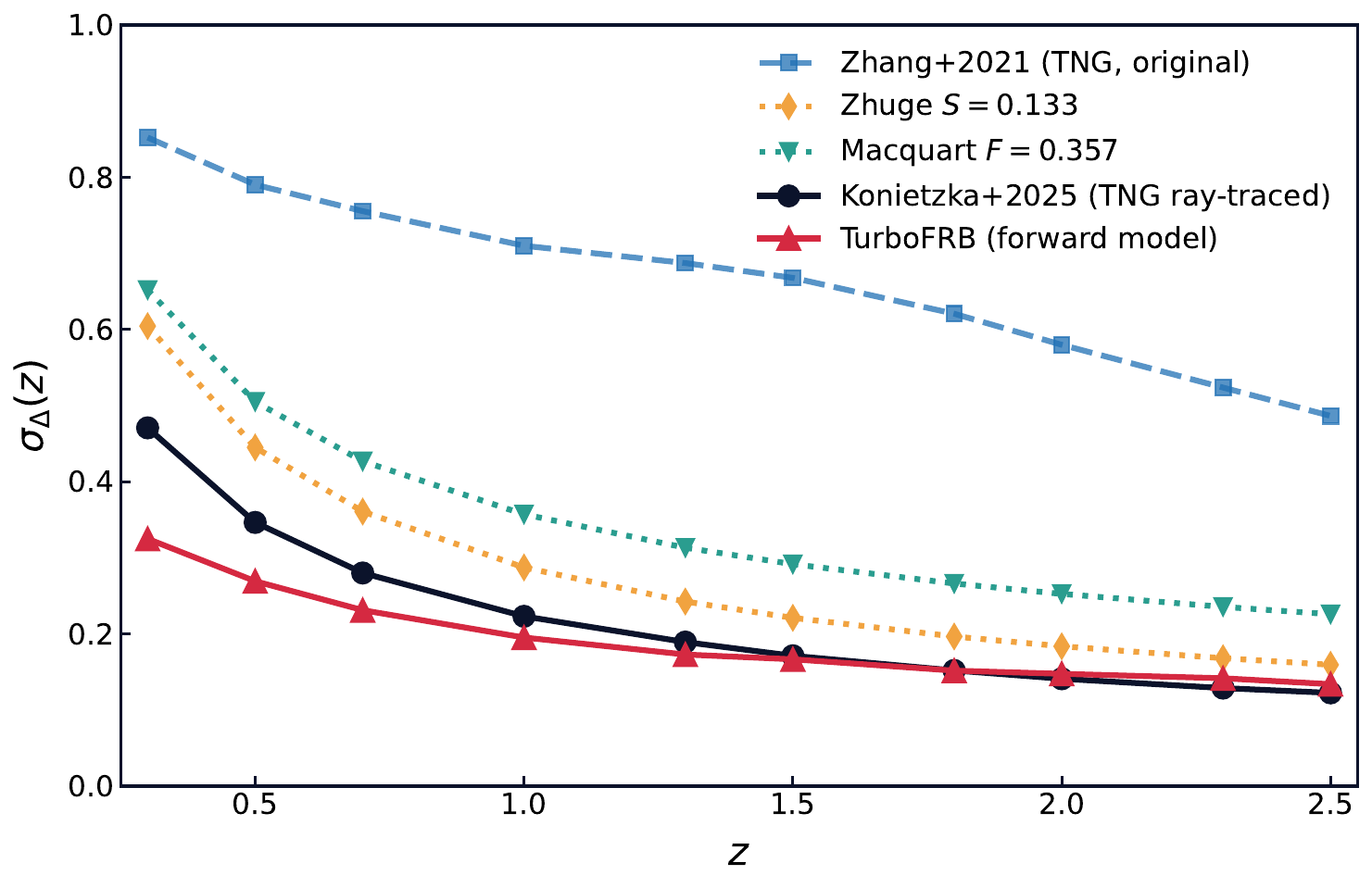}
\caption{Normalised dispersion-measure scatter $\sigma_\Delta(z)$ from \turbofrb{} forward samples (red triangles) compared with the corrected ray-traced TNG measurements of ~\citet{Konietzka2025} (black circles), the Macquart parameterisation with $F=0.357$ (teal inverted triangles), the Zhuge parameterisation with $S=0.133$ (gold diamonds), and the original fit of \citet{Zhang2021} (grey squares). The calibrated \turbofrb{} curve crosses Konietzka near $z\sim 1.5$--$1.8$ and tracks it to within $\lesssim 15\%$ for $z\gtrsim 1$; the deviation grows toward low redshift, where the absolute scatter is largest. No systematic high-redshift floor is present. Note that the \turbofrb{} and Konietzka points are directly measured $\sigma_\Delta$, whereas the Macquart and Zhuge curves are the width parameter $\sigma_{\rm diff}$ of their respective parametric families; their separation at low redshift is the $\sigma_{\rm diff}$--$\sigma_\Delta$ gap measured directly.}
\label{fig:fig2_sigma_delta}
\end{figure}

The most compact summary of the normalised width of the forward distribution is
\be
\sigma_\Delta(z)\equiv
\sqrt{\mathrm{Var}\!\left[\frac{\dmigm}{\avg{\dmigm(z)}}\right]}.
\label{eq:sigma_delta}
\ee
Because \turbofrb{} produces full samples rather than an assumed PDF family, this quantity can be measured directly from the forward realisations without imposing any parametric form, and without the asymptotic approximations that limit the validity of $\sigma_{\rm diff}=F/\sqrt{z}$ at low $z$. This is a conceptual distinction worth making explicit: $\sigma_{\rm diff}$ is the internal width parameter of the Macquart-style PDF family of Eq.~\eqref{eq:usual_pdf}, whereas $\sigma_\Delta$ is the true standard deviation of the normalised variable $\Delta$, and the two are identified only in the asymptotic regime where the parametric form is exact~\citep{Zhuge2026}. \turbofrb{} measures $\sigma_\Delta$ directly and alongside the parametric $\sigma_{\rm diff}$ references within the same construction, so the gap between them ceases to be a formal caveat and becomes a quantity we can read off Fig.~\ref{fig:fig2_sigma_delta} directly. We compare it with three external references: the corrected TNG measurements of ~\citet{Konietzka2025}, the Macquart parameterisation with $F=0.357$~\citep{Macquart2020}, and the Zhuge parameterisation with $S=0.133$~\citep{Zhuge2026}. For historical context, we also show the original Zhang curve \citep{Zhang2021}. Figure~\ref{fig:fig2_sigma_delta} summarises the result.

The most important qualitative feature is that the \turbofrb{} curve \emph{crosses} the Konietzka benchmark within the calibration window rather than lying systematically above it. The two curves are essentially coincident near $z\sim 1.5$--$1.8$, where the $\Delta$-space JSD also attains its minimum. Below the crossing point, the forward model is narrower than Konietzka; above it, the forward model is marginally broader. The agreement is closest at intermediate-to-high redshift: for $z\gtrsim 1$ the relative deviation between the two curves stays within $\lesssim 15\%$, falling to near-coincidence around $z\sim 1.5$--$1.8$ and rising only to $\simeq 10\%$ by $z=2.5$. At the low-redshift end, the deviation is larger --- \turbofrb{} predicts $\sigma_\Delta\simeq 0.33$ at $z=0.3$ against $\simeq 0.47$ for Konietzka, a $\sim 30\%$ difference --- so the close tracking is a statement about the $z\gtrsim 1$ regime rather than the full interval. We note that this low-redshift regime is also where the absolute scatter is largest and where the Macquart $F/\sqrt{z}$ form itself becomes least reliable, and that the great majority of the localised-FRB sample relevant to $P(\dmigm\mid z)$ validation lies at $z\gtrsim 0.3$ with increasing weight toward $z\sim 1$.

\section{Effective halo statistics from per-sightline traces}
\label{sec:halostats}

One advantage of a forward model built from discrete structural encounters is that it can be interrogated at a level deeper than the final PDF \citep{Alfradique:2024fkb}. Most analytic descriptions of the FRB DM scatter compress the effect of halo encounters into a single effective parameter whose value is then quoted as a constant. \turbofrb{} produces the per-sightline encounter catalogue directly, and we can therefore measure each of these quantities as a function of redshift rather than assuming them constant. The resulting redshift dependence is itself the physical content extracted from the forward traces.

For each forward realisation we record the mass, redshift, and impact parameter of every halo intersected along the line of sight. From this set of encounters we define the halo-induced scatter per encounter,
\be
\sigma_{\rm halo}^2(z)=\mathrm{Var}_{\rm enc}[\dmhalo],
\label{eq:sigma_halo}
\ee
the mean number of halo encounters per sightline, $\avg{N_h(z)}$, and the empirical  \citet{Zhuge2026} parameter
\be
S_{\rm emp}(z)=\frac{\sigma_\Delta^2(z)\,J^2(z)}{I(z)},
\label{eq:S_emp}
\ee
which is the value of $S$ that would reproduce the forward-measured variance if inserted into the Zhuge ansatz at that redshift. Because $\sigma_\Delta$ aggregates the diffuse, halo, filament and environmental contributions, $S_{\rm emp}$ is an effective parameter reproducing the \emph{total} normalised variance within that ansatz, not an isolated measurement of the halo channel. Here $I(z)=\int_0^z dz'/E(z')$ and $J(z)=\int_0^z(1+z')\,dz'/E(z')$. We additionally compute
\be
l(z)=\frac{D_c(z)}{\avg{N_h(z)}},
\label{eq:l_eff}
\ee
the effective path length between halo encounters. No DM weighting enters Eq.~\eqref{eq:l_eff}: it is the comoving distance divided by the mean encounter count.

Table~\ref{tab:halo_stats} shows how the 4-parameter calibration shapes the halo channel. The mean number of halo encounters per LoS rises from $\avg{N_h}\simeq 0.47$ at $z=0.2$ to $\simeq 2.9$ at $z=2.5$, while the halo-induced scatter per encounter increases from roughly $49\,\mathrm{pc\,cm}^{-3}$ at low redshift to about $123\,\mathrm{pc\,cm}^{-3}$ by $z=2.5$, and the mean DM contribution per halo encounter grows from $\simeq 37$ to $\simeq 85$~pc\,cm$^{-3}$ over the same interval. The effective encounter path length stabilises at $l\simeq 1.8$--$2.1$~Gpc over the full interval shown. The quantity $l(z)\simeq 2$~Gpc should not be read as the physical halo spacing: it is the path length between DM-relevant halo intersections, that is, those at impact parameters where the projected gas column is non-negligible.

\begin{table}
\centering
\caption{Effective halo statistics extracted from \turbofrb{} per-sightline traces. $\sigma_{\rm halo}$ is the halo-induced DM scatter per encounter; $l = D_c(z)/\langle N_h\rangle$ is the empirical halo path length; $\langle N_h\rangle$ is the mean number of halo encounters per LoS; $\sigma_\Delta$ is the empirical normalized DM scatter; $S_{\rm emp}$ is the inverted Zhuge parameter. The Zhuge published value is $S=0.133^{+0.034}_{-0.045}$.}
\label{tab:halo_stats}
\begin{tabular}{cccccc}
\toprule
$z$ & $\sigma_{\rm halo}$ [pc cm$^{-3}$] & $l$ [Gpc] & $\langle N_h\rangle$ & $\sigma_\Delta$ & $S_{\rm emp}$ \\
\midrule
0.20 & 49 & 1.80 & 0.47 & 0.383 & 0.036 \\
0.30 & 55 & 1.81 & 0.68 & 0.329 & 0.041 \\
0.50 & 66 & 1.89 & 1.02 & 0.267 & 0.049 \\
0.70 & 68 & 1.96 & 1.31 & 0.226 & 0.053 \\
1.00 & 78 & 1.91 & 1.78 & 0.197 & 0.063 \\
1.30 & 86 & 2.00 & 2.04 & 0.168 & 0.065 \\
1.50 & 98 & 1.94 & 2.30 & 0.169 & 0.078 \\
2.00 & 113 & 2.03 & 2.62 & 0.155 & 0.096 \\
2.50 & 123 & 2.06 & 2.90 & 0.137 & 0.099 \\
\bottomrule
\end{tabular}
\end{table}

The empirical $S_\mathrm{emp}(z)$ rises monotonically from $\simeq 0.036$ at $z = 0.2$ to $\simeq 0.10$ by $z = 2.5$, approaching the value $S = 0.133^{+0.034}_{-0.045}$ of \citet{Zhuge2026} from below and entering its $1\sigma$ range only at high redshift. This redshift dependence is not captured by the constant-$S$ ansatz; within the \turbofrb framework, the effective variance parameter increases by nearly a factor of three between $z = 0.2$ and $z = 2.5$, consistent with the growing encounter rate shown in Table~\ref{tab:halo_stats}.
\section{Application to localised FRBs}
\label{sec:outliers}

The forward likelihood $P(\dmigm\mid z)$ is more than a population-level diagnostic: at the level of individual events it provides a quantitative answer to the question of how unusual an observed DM is under the full forward model of Eq.~\eqref{eq:dm_total_conv}, which includes the IGM together with the adopted host and Milky Way priors. Localised FRBs whose observed DM lies far in the tail of the cosmological distribution are a heterogeneous population. Some are produced by genuine cosmological sightlines that happen to sample the high-percentile tail of the variance budget; others are systematically high because they are embedded in dense host environments that contribute a large in-source DM not captured by the cosmological channel. Distinguishing between these possibilities at the level of an individual event has historically been difficult, because the intrinsic cosmological scatter at typical localised-FRB redshifts is large enough to accommodate moderate excess without obvious tension. Our forward likelihood quantifies this directly through the tail probability of each event.

For each localised event, we build the forward distribution of the \emph{total} observable dispersion measure at the host redshift by convolving the three independent contributions of Eq.~\eqref{eq:dm_decomp},
\be
P(\dmtot \mid z_{\rm host}) = P(\dmigm \mid z_{\rm host}) \otimes P(\dmhost^{\rm obs}) \otimes P(\dmmw),
\label{eq:dm_total_conv}
\ee
where $P(\dmigm \mid z_{\rm host})$ is the forward-model output, $P(\dmhost^{\rm obs})$ is the observer-frame host prior defined below in Eq.~\eqref{eq:host_prior}, and $P(\dmmw)$ is a Gaussian centred on the Galactic-foreground estimate. The tail probability is then
\be
P_{\rm tail} = \int_{\dmobs}^{\infty} P(\dmtot \mid z_{\rm host})\,d\dmtot,
\label{eq:tail_prob}
\ee
the probability under the full forward model --- cosmological IGM plus the assumed host and Milky Way priors --- of obtaining a total DM at least as large as the observed value. A small $P_{\rm tail}$ therefore means the observation is unlikely even after the typical host and foreground contributions have been folded in, which is the quantitative signature of a required host excess. We adopt the convention $\dmmw \simeq 30$~pc\,cm$^{-3}$ for the Galactic halo contribution following the recent estimate of ~\citet{Connor2023}; we note that higher values up to $50$--$80$~pc\,cm$^{-3}$ have been favoured by ~\citet{ProchaskaZheng2019} on the basis of hydrostatic CGM modelling, and that adopting the higher range would shift the inferred $\dmigm^{\rm obs}$ of each event downward by tens of pc\,cm$^{-3}$ without changing the qualitative reading of any event in the sample below.

For the host contribution, we adopt a lognormal prior, the standard choice in the FRB literature~\citep{Macquart2020}, with rest-frame parameters measured by ~\citet{SangLin2025} from a sample of 117 localised FRBs,
\be
P(\dmhost^{\rm rest}) = \frac{1}{\dmhost^{\rm rest}\,\sigma_{\rm host}\sqrt{2\pi}}\exp\!\left[-\frac{\left(\ln \dmhost^{\rm rest} - \mu_{\rm host}\right)^2}{2\sigma_{\rm host}^2}\right],
\label{eq:host_prior}
\ee
with $\mu_{\rm host}=5.03$ and $\sigma_{\rm host}=0.96$ (median $\simeq 153$~pc\,cm$^{-3}$), transformed to the observer frame through $\mu_{\rm host}\to\mu_{\rm host}-\ln(1+z_{\rm host})$. \citet{SangLin2025} also report a positive correlation between the rest-frame host DM and redshift, which we do not model: we adopt a redshift-independent rest-frame prior and apply only the $(1+z)^{-1}$ dilution. An evolving prior would raise the predicted host contribution at high redshift and hence increase $P_{\rm tail}$ there, so the values quoted for FRB\,20240304B and FRB\,20210912A understate, if anything, how typical those events are under an evolving-host model.

\subsection{Tail-probability criteria for interpreting individual events}
\label{ssec:classification}

We use $P_{\rm tail}$ only as a descriptive statistic, not as a model-selection criterion. To make the event-level discussion reproducible, we group events into four regimes:
\begin{center}
\begin{minipage}{0.68\linewidth}
\[
\begin{aligned}
P_{\rm tail}<1\% &: \quad \text{host excess required},\\
1\%\leq P_{\rm tail}<5\% &: \quad \text{moderate host excess},\\
5\%\leq P_{\rm tail}<20\% &: \quad \text{cosmic-variance tail},\\
P_{\rm tail}\geq20\% &: \quad \text{typical sightline}.
\end{aligned}
\]
\end{minipage}
\end{center}
These boundaries are pragmatic and can be changed without altering the likelihood itself. The physical interpretation of any individual event should therefore be based on the reported $P_{\rm tail}$ value together with independent host and foreground information.
\subsection{Representative events}
\label{ssec:events}

\begin{figure*}
\centering
\includegraphics[width=0.92\textwidth]{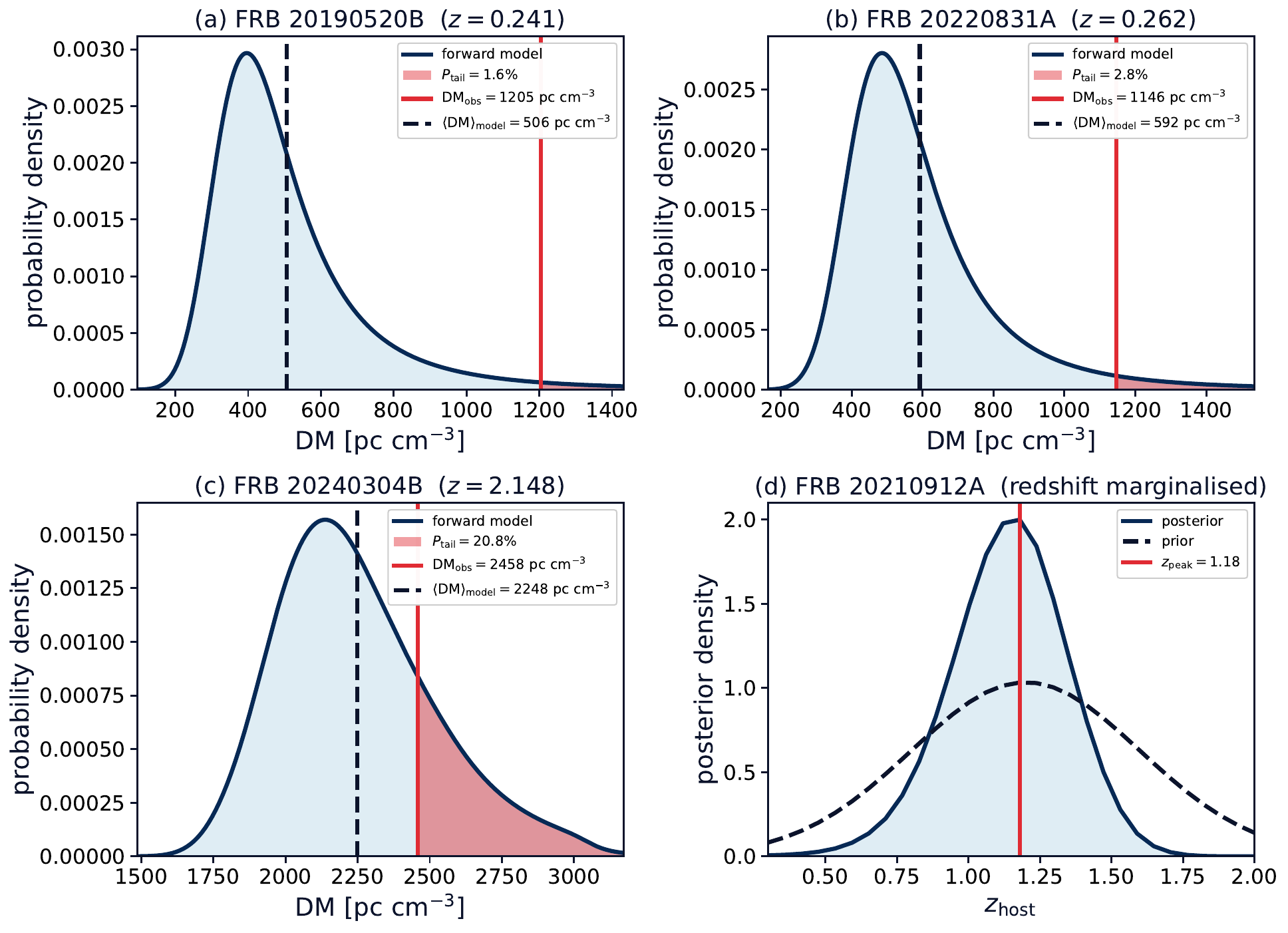}
\caption{Forward-model diagnostic for the four events of Section~\ref{ssec:events} under the frozen 4-parameter \turbofrb{} configuration. Panels (a)--(c) show the forward distribution of the total observable DM at the host redshift (blue filled curve) with the observed $\dmobs$ marked (red solid vertical line) and the forward-model mean $\avg{\dmtot}_{\rm model}$ marked (dark dashed vertical line); the tail above $\dmobs$ is shaded in red. The tail probability $P_{\rm tail}$, $\dmobs$, and $\avg{\dmtot}_{\rm model}$ are reported in the legend of each panel. Panel (d) shows the posterior on $z_{\rm host}$ for the host-undetected event FRB\,20210912A, marginalising the forward likelihood over a Gaussian prior $\mathcal{N}(\mu=1.2,\sigma=0.4)$ on the host redshift.}
\label{fig:fig4_outliers}
\end{figure*}

We apply the diagnostic to four events in total: three with a confirmed host redshift, spanning $z=0.241$ to $z=2.148$, and one localised in radio but without an identified host, whose redshift is marginalised over. Two of the three localised events fall in the host-excess regimes defined above; the third, FRB\,20240304B, is the highest-redshift localised event currently known and probes the high-redshift end of the calibration window. The host-undetected event illustrates the use of the forward likelihood for inferring $z_{\rm host}$ when no spectroscopic confirmation is available. The four panels are shown in Fig.~\ref{fig:fig4_outliers}, and the corresponding numerical values are tabulated in Table~\ref{tab:outliers}.

\medskip
\noindent\textbf{FRB\,20190520B ($z=0.241$):}
With $\dmobs=1204.7$~pc\,cm$^{-3}$ and assumed Galactic foregrounds of $\dmmw\simeq 90$~pc\,cm$^{-3}$, this event lies deep in the tail of the cosmological forward distribution at its host redshift: $P_{\rm tail}\simeq 1.6\%$, against a forward-model mean $\avg{\dmtot}_{\rm model}\simeq 506$~pc\,cm$^{-3}$. In our classification, this event sits at the boundary of the \emph{host-excess required} regime. The reading is consistent with the independent observational characterisation of FRB\,20190520B as a persistent host-excess source on the basis of its compact persistent radio source~\citep{Niu2022}, its rapidly varying Faraday rotation measure~\citep{Anna-Thomas2023}, and its dense local environment~\citep{Niu2022}. The residual required to reconcile the observation with the forward model is large; we quantify it explicitly in Section~\ref{ssec:host_posteriors}. Spectroscopy of this field has since identified two foreground galaxy clusters with $M_{\rm halo}>10^{14}\,M_\odot$, at $z=0.1867$ and $z=0.2170$, whose characteristic radii are directly intersected by the sightline~\citep{Lee2023}, and which account for a substantial part of the excess previously attributed to the host. The forward likelihood therefore correctly flags this sightline as anomalous, but does not by itself localise the origin of the excess: the residual of Section~\ref{ssec:host_posteriors} is the total unmodelled line-of-sight excess, foreground structures plus host, and is an upper bound on the host term.

\noindent\textbf{FRB\,20220831A ($z=0.262$):}
With $\dmobs=1146.2$~pc\,cm$^{-3}$ and $\dmmw\simeq 157$~pc\,cm$^{-3}$, this event has $P_{\rm tail}\simeq 2.8\%$ under the full forward model, including the host and Milky Way priors, against a forward-model mean $\avg{\dmtot}_{\rm model}\simeq 592$~pc\,cm$^{-3}$. In our classification, this event sits in the \emph{moderate host-excess} regime: the DM is uncommon but not extreme, and the required host contribution is moderately smaller than for FRB\,20190520B. The fact that two qualitatively distinct events (one with a confirmed dense local environment, one without similarly direct spectroscopic evidence) both fall in the $P_{\rm tail}<5\%$ region of the diagnostic confirms that the forward likelihood is sensitive to a real cosmological-versus-host discrepancy, and not to the specific astrophysical signatures that motivated each event independently.

\noindent\textbf{FRB\,20240304B ($z=2.148$):}
The recent MeerKAT detection FRB\,20240304B at $z=2.148$~\citep{Caleb2025} is currently the highest-redshift localised FRB. With $\dmobs=2458.2$~pc\,cm$^{-3}$ and $\dmmw\simeq 58$~pc\,cm$^{-3}$ from the NE2001 model~\citep{cordes2002ne2001} plus a halo term, the event has $P_{\rm tail}\simeq 20.8\%$ under the calibrated \turbofrb{} forward model, against a forward-model mean $\avg{\dmtot}_{\rm model}\simeq 2248$~pc\,cm$^{-3}$. The event therefore falls just inside the \emph{typical sightline} regime defined above: the observed DM is consistent with the body of the cosmological distribution at this redshift, with no host or foreground excess strictly required, though it sits close to the cosmic-variance-tail boundary and a modest host contribution is equally compatible. Two cautions are worth stating. First, even though the event is statistically unremarkable in the forward model, its high $\dmigm^{\rm obs}$ probes the calibrated $P(\dmigm\mid z)$ at the high-redshift end of the calibration window, where the corrected hydrodynamical reference is sparsest, since the Konietzka catalogue at $z>2$ is populated by relatively few simulation snapshots. Second, with a single $z>2$ event we cannot empirically discriminate between the broader \turbofrb{} prediction and the narrower Konietzka prediction at $z\gtrsim 2$, despite the fact that the Konietzka tail probability for the same event would be smaller. A population of localised $z>2$ FRBs will be required to test whether the empirical tail at high redshift is closer to the forward-model prediction or to the corrected hydrodynamical reference.

\noindent\textbf{FRB\,20210912A (host-undetected):} For events without a spectroscopically confirmed host redshift, the same forward likelihood can be used in reverse. Instead of evaluating $P_{\rm tail}$ at a known $z_{\rm host}$, we combine the observed DM with a redshift prior to infer a posterior on $z_{\rm host}$. FRB\,20210912A is a useful stress test for this procedure: although it was localised to sub-arcsecond precision, no host galaxy was identified in deep optical and near-infrared follow-up \citep{marnoch2023unseen}. Its large observed dispersion measure, $\dmobs=1233.7~{\rm pc\,cm^{-3}}$, suggests a high-redshift origin unless a substantial fraction of the excess DM is produced by the host or local environment. Using our adopted Galactic contribution, $\dmmw\simeq61~{\rm pc\,cm^{-3}}$, and a broad Gaussian prior $p(z_{\rm host})=\mathcal{N}(\mu=1.2,\sigma=0.4)$ centred on the photometric redshift range favoured by \citet{marnoch2023unseen} for the field, the posterior peaks at $z_{\rm peak}\simeq1.18$. The peak therefore largely reflects the prior; the DM likelihood alone is broad and only excludes $z\lesssim0.6$ without a large host term. The result is therefore consistent with the interpretation that FRB\,20210912A lies at $z\sim1.2$, while still allowing lower-redshift solutions if the host or local DM contribution is unusually large. The posterior is shown in panel~(d) of Fig.~\ref{fig:fig4_outliers}. This example illustrates how the same forward machinery used for event-level diagnostics can also provide a likelihood-based redshift estimator for FRBs lacking a spectroscopic host redshift.

\begin{table*}
\centering
\caption{Forward-likelihood diagnostics for the FRBs analysed in Section~\ref{sec:outliers}. $\langle\mathrm{DM}\rangle_{\rm model}$ is the mean total DM predicted by \texttt{turboFRB} under the frozen 4-parameter calibration with host and Milky-Way priors; $P_{\rm tail}\equiv P(\mathrm{DM}_{\rm tot}\geq\mathrm{DM}_{\rm obs})$ under the same forward distribution.}
\label{tab:outliers}
\begin{tabular}{lcccc}
\toprule
FRB & $z_{\rm host}$ & DM$_{\rm obs}$ [pc cm$^{-3}$] & $\langle$DM$\rangle_{\rm model}$ [pc cm$^{-3}$] & $P_{\rm tail}$ \\
\midrule
FRB 20190520B & 0.241 & 1204.7 & 506 & 1.6\% \\
FRB 20220831A & 0.262 & 1146.2 & 592 & 2.8\% \\
FRB 20240304B & 2.148 & 2458.2 & 2248 & 20.8\% \\
FRB 20210912A & $\sim 1.18$ (post.) & 1233.7 & --- & --- \\
\bottomrule
\end{tabular}
\end{table*}

\subsection{Implied host dispersion measure for the localised events}
\label{ssec:host_posteriors}

The tail probability answers ``how unusual is this event?'', but for the events that do require a host excess it is more physically informative to ask ``how much host DM would reconcile the observation with the cosmological forward model?''. We answer this by inverting the decomposition~\eqref{eq:dm_decomp} stochastically: for each forward draw of $\dmigm$ from $P(\dmigm\mid z_{\rm host})$, we solve for the rest-frame host contribution
\be
\dmhost^{\rm rest} = (1+z_{\rm host})\,\bigl(\dmobs - \dmmw - \dmigm\bigr),
\label{eq:host_inversion}
\ee
which converts the tail of the forward distribution into the implied distribution of the rest-frame residual $\dmhost^{\rm rest}$. Because no explicit host prior enters this inversion, and negative values are retained for bookkeeping, this is a required-residual distribution rather than a Bayesian posterior on the host DM. Draws for which the sampled IGM already exceeds the available budget yield $\dmhost^{\rm rest}\le 0$ and are reported as the fraction $f_{\rm no\,host}$ requiring no residual. The resulting distributions for the three localised events are summarised in Table~\ref{tab:host_posteriors}.

\begin{table}
\centering
\caption{Implied rest-frame host dispersion measure required to reconcile each localised outlier with the cosmological forward model, obtained by inverting $\mathrm{DM}_{\rm host}^{\rm rest} = (1+z)(\mathrm{DM}_{\rm obs}-\mathrm{DM}_{\rm MW}-\mathrm{DM}_{\rm IGM})$ with $\mathrm{DM}_{\rm IGM}$ sampled from the forward distribution at $z_{\rm host}$. Values are the median and 68\% credible interval; $f_{\rm no\,host}$ is the fraction of forward draws for which the IGM alone already accounts for the observed budget.}
\label{tab:host_posteriors}
\begin{tabular}{lccc}
\toprule
FRB & $z_{\rm host}$ & $\mathrm{DM}_{\rm host}^{\rm rest}$ [pc cm$^{-3}$] & $f_{\rm no\,host}$ \\
\midrule
FRB 20190520B & 0.241 & $1115^{+69}_{-98}$ & 0.00 \\
FRB 20220831A & 0.262 & $950^{+76}_{-109}$ & 0.00 \\
FRB 20240304B & 2.148 & $989^{+660}_{-1021}$ & 0.17 \\
\bottomrule
\end{tabular}
\end{table}

For FRB\,20190520B the inferred host contribution is $\dmhost^{\rm rest}=1115^{+69}_{-98}$~pc\,cm$^{-3}$ with no draws in the forward ensemble requiring zero residual ($f_{\rm no\,host}$ consistent with zero at the resolution of the sample): the required residual is well above the median of any standard lognormal host prior. Given the foreground clusters along this sightline~\citep{Lee2023}, it should be read as the sum of the host term and the excess from the intersected structures rather than as a host measurement; the dense magneto-ionic local environment inferred for this source~\citep{Niu2022,Anna-Thomas2023} independently supports a substantial, though not necessarily dominant, host contribution. FRB\,20220831A requires a smaller but still significant residual, $\dmhost^{\rm rest}=950^{+76}_{-109}$~pc\,cm$^{-3}$, again with no zero-residual draws observed. FRB\,20240304B, by contrast, has a broad posterior consistent with zero, $\dmhost^{\rm rest}=989^{+660}_{-1021}$~pc\,cm$^{-3}$ with $f_{\rm no\,host}=0.17$: in $17\%$ of forward draws the cosmological IGM alone already accounts for the observed budget before any host contribution; folding in the host and Milky-Way priors gives the tail probability $P_{\rm tail}\simeq 21\%$, consistent with the requirement $P_{\rm tail}\geq f_{\rm no\,host}$. This event therefore does not require a host excess, but does not exclude a moderate one either.

\section{Brief proof-of-concept for $H_0$ inference}
\label{sec:h0}

The same forward likelihood can be embedded in a cosmological analysis because both $\avg{\dmigm(z)}$ and $P(\dmigm\mid z,H_0)$ depend on the expansion history. As a closed-loop consistency check, we generate a synthetic catalogue of $N=100$ localised FRBs from the frozen 4-parameter model with $H_0^{\rm true}=67.74\,$km\,s$^{-1}$\,Mpc$^{-1}$, add host and Milky-Way contributions from the same priors used in the likelihood, and recover $H_0$ on a one-dimensional grid. The posterior is centred on the injected value and contains it within $1\sigma$. Because the injection and recovery use the same calibration, host prior, Milky Way prior, and $f_{\rm IGM}$ prescription, this test should not be read as a forecast. It only verifies that the likelihood implementation is internally consistent; a real-data analysis with nuisance-parameter marginalisation is deferred to future work.

\section{Discussion}
\label{sec:discussion}

Different and independent methodologies now converge on a common scale for the FRB DM scatter: the corrected TNG ray-tracing of \cite{Konietzka2025}, the analytic baryonification model of \cite{Torkamani2026}, the empirical FRB-data parameterisations of \cite{Macquart2020} and \cite{Zhuge2026}, and the present forward model. We keep this statement qualitative: the references do not coincide exactly at every redshift, and \turbofrb{} retains a residual shape mismatch at the edges of the window (a marginally heavier high-DM tail at $z=2.5$ and a narrower body at the lowest redshifts).

It is worth situating \turbofrb{} relative to the two methodologically closest recent works, both of which target the same DM distribution but by different routes. \cite{Konietzka2025} decompose the cosmic-web DM directly in IllustrisTNG by classifying each sightline segment as halo, filament, or void in post-processing of the simulation; this is a measurement of the simulated decomposition instead of a parametric generator, and it is complementary to our approach in that it provides the physical target our forward channels are designed to reproduce. \cite{Torkamani2026}, by contrast, construct a fully analytic DM PDF by convolving halo contributions weighted by the halo mass function and halo bias within the baryonification framework; this is an analytic halo-model calculation rather than a stochastic forward sampler, and in its present form it does not include an explicit geometrically distinct filament channel. \turbofrb{} occupies the intermediate position of a stochastic forward generator: like the analytic models it is fast and parametric, but like the ray-tracing it produces full per-sightline realisations from which arbitrary statistics can be measured directly rather than assumed.

A useful external cross-check on the calibrated halo channel is provided by the same cosmic-web decomposition. \citet{Walker2024} report that an FRB sightline originating at $z=1$ intersects on average $\simeq 1.8$ foreground collapsed structures in TNG, rising to $\simeq 12.4$ by $z=5$. Our calibrated mean halo-encounter count at the same redshift is $\avg{N_h}\simeq 1.8$ (Table~\ref{tab:halo_stats}), of the same order. That count includes collapsed structures down to subhalo masses $M\sim 10^{8}\,h^{-1}M_\odot$, whereas our halo channel samples the more massive range $M\in[10^{12},10^{14.8}]\,M_\odot$ that dominates the projected gas column, so the two counts are not defined over identical populations. What is nonetheless reassuring is that the calibration, which had no access to the encounter count and was driven purely by the $\Delta$-space shape, lands on an encounter rate of the same order as the structure-intersection rate measured directly in the simulation.

\subsection{Filaments as a physical channel}

A separate but complementary line of evidence supports the explicit treatment of filaments as a contribution to $\dmigm$. \cite{ConnorFil2025} have detected at $\sim 3\sigma$ confidence a divergence in the DM-redshift relation between FRBs whose lines of sight intersect cosmic filaments identified by the DisPerSE algorithm in DESI imaging, and FRBs whose lines of sight do not, with the modelled filament gas described by an isothermal $\beta$-model with central baryon overdensity $\delta_0 = 21^{+13}_{-12}$. The filament channel of \turbofrb{} adopts the same isothermal $\beta$-model functional form, while its absolute amplitude is fixed by budget enforcement. The channel decomposition of Fig.~\ref{fig:figB_decomp} shows that, at $z=1$, the filament channel contributes a mean of $\simeq 51$~pc\,cm$^{-3}$ but with a sharply skewed distribution, consistent with its expected role in populating the high-DM shoulder rather than the body.

The explicit filament channel remains one of the methodological commitments that distinguishes \turbofrb{} from purely halo-based forward models. Within the present calibration strategy, the main improvement did not come from changing the filament sector, but from altering the halo encounter normalisation. That itself is informative: it suggests that the halo channel primarily controlled the residual variance budget, while the filament channel plays a secondary but still physically motivated role in shaping the shoulder and high-DM tail of $P(\dmigm\mid z)$.

\subsection{Implications for FRB cosmology}

The recent demonstration by \cite{AndrewMasui} that the FRB dispersion measure is an effectively unbiased tracer of large-scale matter, with a bias parameter $b_{\rm DM} \simeq 1$ on linear scales, has important implications for the cosmological use of FRBs. This result was derived under the assumption that the DM field traces the matter distribution without resolving the contributions of individual structures. Forward models such as \turbofrb{} that decompose the DM into explicit halo and filament channels are therefore natural tools for testing and extending this prediction, since halos and filaments are known to be biased tracers of the matter field with different bias parameters. They are particularly suited for cross-correlations between FRB DMs and galaxy positions, between FRB DMs and weak-lensing maps, and between FRB DMs at different sightline pairs. The present paper does not attempt those analyses, but it establishes the forward machinery required for them.

A second implication concerns the use of the DM-redshift relation for cosmological parameter inference, particularly the Hubble constant $H_0$. The forward model exposes the $H_0$ dependence of $\avg{\dmigm}$ but also provides the full $P(\dmigm\mid z,H_0)$ likelihood, allowing for Bayesian inference that does not rely on a Gaussian approximation around the mean relation. The closed-loop test of Section~\ref{sec:h0} shows that the inference machinery built on this likelihood recovers the injected value when fed mock data drawn from the same generative model; a realistic analysis on real localised FRBs, including the joint marginalisation of $f_{\rm IGM}$, the host prior parameters, and the four \turbofrb{} calibration parameters, is left to a forthcoming companion paper.

\subsection{Limitations and future work}
\label{ssec:limitations}

First, the host-galaxy contribution $\dmhost$ is not modelled self-consistently; in Section~\ref{sec:outliers} it is treated as a residual diagnosed through the forward likelihood, and in Section~\ref{sec:h0} it is encoded through a fixed lognormal prior. A complete cosmological likelihood will require a joint host+IGM treatment, ideally along the lines of recent host-galaxy forward models~\citep{Theis2024}.

Second, a residual shape mismatch remains between the calibrated forward distribution and the corrected TNG reference at the edges of the calibration window: the calibrated $\sigma_\Delta(z)$ crosses the Konietzka benchmark near $z\sim 1.5$--$1.8$ and tracks it to within $\lesssim 15\%$ for $z\gtrsim 1$, but the deviation grows toward low redshift (reaching $\sim 30\%$ at $z=0.3$), and a marginally heavier high-DM tail persists at $z=2.5$. We attribute this residual to the cumulative effect of the engine's remaining fixed assumptions (Tinker mass function, halo concentration, filament cylinder geometry, $f_{\rm gas}$ values, environmental coupling exponents), each of which is anchored at a literature value rather than calibrated. Releasing any one of these to a calibrated effective parameter would almost certainly reduce the residual further, at the cost of moving the model away from its currently transparent physical interpretation. We regard the present four-parameter configuration as a reasonable compromise between calibration quality and physical structure.

Third, the environmental modulation as implemented here is a simplification of the full cosmic-web structure: the latent $E$ is drawn independently per line of sight, with no explicit angular correlation, while in nature the environmental overdensity field has substantial coherence on Mpc scales. This limitation is irrelevant for the marginal $P(\dmigm\mid z)$ that is the focus of the present paper, but it becomes relevant for FRB cross-correlation analyses, which we leave to future work.

Fourth, the engine currently propagates only the background cosmological sector. The diffuse mean~\eqref{eq:dm_mean_analytic} depends on $H_0$, $\Omega_m$, and $\Omega_b$, and the encounter rates inherit the same background dependence; but perturbation-sector parameters such as $\sigma_8$ and $n_s$, which would in principle modulate the halo mass function and therefore $n_{\rm hbg}$, are not coupled dynamically and instead get absorbed into the calibrated effective $n_{\rm hbg}$ and into the fixed log-normal pool parameters. For the present study, in which the calibration is performed against a single hydrodynamical benchmark at fixed and matched cosmology, this is acceptable. For the broader cosmological applications, the engine will be extended to accept an external cosmology object (for instance via \textsc{astropy.cosmology}, \textsc{CAMB}, or \textsc{CLASS}) so that the forward likelihood can be embedded in a sampler such as \textsc{Cobaya} or \textsc{CosmoSIS} and used to constrain background cosmology jointly with the host-galaxy and \turbofrb{} nuisance parameters. Such a coupling would also permit a controlled study of how $\sigma_8$ and $n_s$ propagate into the FRB DM distribution through the halo channel, which the present engine cannot address.

\section{Conclusions}
\label{sec:conclusions}

We have presented \turbofrb{}, a semi-analytic stochastic generator for FRB intergalactic dispersion measures. The model builds $P(\dmigm\mid z)$ from a mean-preserving diffuse component, Poisson-sampled halo and filament encounters, and a latent line-of-sight environmental modulation. Our main results are:

\begin{enumerate}[leftmargin=*]
\item Of the 15 scalar parameters exposed by the sampling configuration, four are calibrated as effective nuisance parameters by a single-shot grid search against the \citet{Konietzka2025} ray-traced TNG catalogue in $\Delta$-space, at matched (TNG) cosmology: $\sigma_0=0.375$, $\gamma_s=-1.460$, $\sigma_E=0.450$, and $n_{\rm hbg}=2.60\times 10^{-3}\,\mathrm{Mpc}^{-3}$. The remaining parameters are fixed at literature values, engine defaults, or budget-closure relations.

\item The calibrated forward model matches the corrected TNG benchmark to the percent level in mean DM across the calibration window (mean ratios $0.982$--$0.990$) and yields a per-redshift $\Delta$-space Jensen--Shannon divergence~\citep{lin1991divergence} at or below $\simeq 5\times 10^{-3}$ across $z=0.5$ to $z=2.5$. Calibrating at the benchmark cosmology ensures the four effective parameters capture genuine shape differences rather than a cosmological-parameter offset.

\item The directly measured normalised scatter $\sigma_\Delta(z)$ \emph{crosses}  the \citet{Konietzka2025} benchmark near $z\sim 1.5$--$1.8$, rather than lying systematically above it, and tracks it to within $\lesssim 15\%$ for $z\gtrsim 1$; the deviation grows toward low redshift, where the absolute scatter is largest. The high-redshift plateau in $\sigma_\Delta$ characteristic of a 3-parameter baseline with $n_{\rm hbg}$ frozen at the Tinker integral is absent in the present calibration.

\item The per-sightline halo statistics provide a physically interpretable effective decomposition of the shape match. The halo encounter population becomes more frequent, with $\avg{N_h}$ rising from $\simeq 0.47$ at $z=0.2$ to $\simeq 2.9$ by $z=2.5$, while the per-encounter halo scatter increases only moderately, from $\simeq 49$ to $\simeq 123$~pc\,cm$^{-3}$. The effective encounter path length stabilises near $l\simeq 2.0$~Gpc.

\item The event-level likelihood framework is applied to four representative FRBs: three with confirmed host redshifts spanning $z=0.241$--$2.148$, and one radio-localised, host-undetected event whose redshift is marginalised over. The tail probability $P_{\rm tail}$ provides a single quantitative answer to ``how unusual is this event under the cosmological forward model?'', and inverting the forward likelihood yields the implied rest-frame residual DM distribution for each event; the interpretive classification (host-excess required, moderate host-excess, cosmic-variance tail, typical sightline) is left to the user as descriptive vocabulary rather than encoded in the engine.

\item A closed-loop $H_0$ self-consistency test on a synthetic catalogue of $N=100$ localised FRBs recovers the injected value within $1\sigma$. The artificially small posterior width is the result of holding the calibration covariance, the $f_{\rm IGM}$ marginalisation, and the host prior uncertainty fixed on both sides of the loop, and should not be read as a realistic forecast; it demonstrates internal consistency of the forward likelihood; a single closed-loop realisation does not establish absence of bias, which would require a coverage study over many realisations.
\end{enumerate}

The present implementation is limited mainly by the absence of a self-consistent host-galaxy DM model and by calibration against a single hydrodynamical benchmark. Incorporating host-galaxy forward modelling, calibration uncertainty, and baryon-fraction marginalisation is the next step toward applying \turbofrb{} to real FRB cosmological inference.

\section*{Acknowledgments}

We thank the authors of~\citet{Konietzka2025} for making their ray-tracing catalogues publicly available. AW and JASF gratefully acknowledge financial support from the South African Research Chairs Initiative (SARChI) of the Department of Science and Technology and the National Research Foundation (NRF) of South Africa, as well as support from the National Institute for Theoretical and Computational Sciences (NITheCS). VM thanks CNPq (Brazil), CAPES (Brazil) and FAPES (Brazil) for partial financial support. WSHR thanks FAPES (Brazil) and CNPq (Brazil) for partial financial support. 

\section*{Data Availability}
 
The code implementing \turbofrb{} is publicly available at
\url{https://github.com/jefersonfortunato/turbofrb}. TNG ray-tracing catalogues used for calibration are publicly released by \citet{Konietzka2025}. The frozen four-parameter configuration, the random seeds, and a script reproducing every figure and table of this paper are included in the repository release corresponding to this paper.


\bibliographystyle{mnras}
\bibliography{turbofrb} 



\appendix

\section{Provenance of fixed inputs and modelling choices}
\label{app:fixed_params}

The calibration described in Section~\ref{ssec:calibration} adjusts only four effective parameters. The remaining quantities entering the forward engine are fixed either by external physics, by representative values from published scaling relations, or by internal modelling choices. We refer to them here as fixed \emph{inputs} rather than fixed parameters, since several entries in Table~\ref{tab:fixed_interp} are grouped modelling choices rather than independent scalar degrees of freedom.

We use three provenance categories: \textbf{(i)} quantities fixed by atomic physics or by direct external constraints; \textbf{(ii)} representative single values adopted from published relations without evolving the full relation dynamically; and \textbf{(iii)} internal engine choices that define the structure of the forward model. The purpose of the table is not to assign a formal prior to each input, but to make clear which parts of the model are calibrated and which are held fixed.

\begin{table*}
\centering
\caption{Provenance and role of the fixed inputs used by \turbofrb{}. Category \textbf{(i)} denotes atomic-physics or direct external constraints; \textbf{(ii)} denotes representative values from published relations; and \textbf{(iii)} denotes internal engine choices.}
\label{tab:fixed_interp}
\begin{tabularx}{\textwidth}{lcl X}
\toprule
Input & Value & Category & Role in the forward model \\
\midrule
$\chi_e$                          & $0.88$                       & (i)   & Electron-per-baryon ratio for fully ionised primordial gas \\
$f_{\rm gas,h}$                & $0.10$                       & (ii)   & Gas fraction assigned to halo encounters~\citep{ProchaskaZheng2019} \\
$\beta$                        & $2/3$                        & (ii)   & Isothermal index of the filament gas profile~\citep{ConnorFil2025} \\
$f_{\rm gas,f}$                & $0.28$                       & (iii) & Filament gas normalisation before budget enforcement \\
$c_{200}$                      & $5.0$                        & (ii)  & Representative NFW concentration for the halo gas profile~\citep{Duffy2008} \\
halo mass pool                 & $10^{12}$ -- $10^{14.8}\,M_\odot$ & (ii)  & Mass range for cross-section-weighted halo encounters~\citep{Tinker2008} \\
$x_{\rm max}$                  & $0.85$                       & (iii) & Maximum halo impact parameter in units of $R_{200}$ \\
$R_f^{\rm ref},L_f^{\rm ref}$  & $0.65,6.0$~Mpc               & (iii) & Reference filament radius and length \\
$R_f(M),L_f(M)$ scalings       & $M^{0.20},M^{0.15}$ ($M_{\rm ref}=10^{13}M_\odot$)          & (iii) & Mass dependence of the finite-cylinder filament geometry \\
spine concentration            & $1.0$                        & (iii) & No additional filament-spine boost in the fiducial model \\
$\eta_h,\eta_f,\eta_{n_h},\eta_{\lambda_f}$ & $\sim\!\pm 0.1$ -- $0.2$ & (iii) & Mild redshift evolution of budget weights and encounter rates \\
$a_h,a_f$                      & $0.60,0.45$                  & (iii) & Response of halo and filament rates to the latent environment \\
halo mass-pool shape           & $\mu_{\log_{10}M}=12.65$, $\sigma_{\log_{10}M}=0.35$ & (iii) & Log-normal pool from which halo encounters are drawn \\
filament mass pool             & $10^{11.8}$--$10^{14.2}\,M_\odot$, $\mu_{\log_{10}M}=12.55$, $\sigma_{\log_{10}M}=0.22$ & (iii) & Log-normal pool for filament encounters \\
gas-fraction scatter           & $0.12$ (halo), $0.10$ (filament) & (iii) & Log-amplitude of per-encounter gas-mass scatter \\
$\epsilon$                     & $0.15$                       & (iii) & NFW core regularisation, in units of $r_s$ \\
$r_c$                          & $0.18\,R_f$                  & (iii) & Filament $\beta$-model core radius \\
$R_{200}^{\rm ref}$            & $0.30$~Mpc at $M_{\rm ref}=10^{13}M_\odot$ & (iii) & Virial-radius normalisation \\
\bottomrule
\end{tabularx}
\end{table*}

Most of the fixed inputs affect the shape of individual channel contributions rather than the ensemble mean, which is controlled by the budget-enforcement step of Section~\ref{ssec:budget}. This separation is useful because changes in profile shape, impact-parameter cuts, or environmental response alter the width and tails of $P(\dmigm\mid z)$ without freely shifting the mean relation. In particular, because the budget step fixes each channel mean and the environmental modulation is mean-preserving by construction (Section~\ref{ssec:environment}), the least externally anchored inputs --- $x_{\rm max}$ and the couplings $a_h,a_f$ --- cannot shift the ensemble mean and act only on the second moment and the tails. None of the fixed inputs was adjusted in light of the Konietzka comparison, which is what makes the residual mismatch (Section~\ref{ssec:sigma_compare}) interpretable as the cumulative rigidity of these choices rather than exhausted fitting freedom. A full sensitivity scan over them is left to future work.

\section{Monte Carlo convergence of the forward estimator}
\label{app:convergence}

All distribution-level quantities reported in this work are estimated from finite ensembles of forward sightlines. We adopt $N=10^4$ sightlines per redshift as the fiducial sample size and test its convergence by recomputing the normalised scatter $\sigma_\Delta(z)$ at $z=0.5,1.5,$ and $2.5$ for
\[
N\in\{2,5,10,20,50\}\times 10^3 .
\]
Each configuration is repeated with five independent random seeds. The results are shown in Table~\ref{tab:convergence}.

Between $N=10^4$ and $N=5\times10^4$, the central value of $\sigma_\Delta$ changes by less than $0.002$ at all three redshifts, smaller than the seed-to-seed scatter at the fiducial sample size. We therefore regard the fiducial ensemble as sufficient for the validation statistics of Section~\ref{sec:validation}; increasing the sample size would reduce Monte Carlo noise but would not change the conclusions of the study.

\begin{table}
\centering
\caption{Convergence of the forward-model normalised scatter $\sigma_\Delta(z)$ with the number of sampled sightlines $N$. Each entry is the mean over 5 independent realisations, with the standard deviation across realisations in parentheses. The estimator is converged at the fiducial $N=10^4$ to better than the per-bin scatter.}
\label{tab:convergence}
\begin{tabular}{lccc}
\toprule
$N$ & $\sigma_\Delta(z=0.5)$ & $\sigma_\Delta(z=1.5)$ & $\sigma_\Delta(z=2.5)$ \\
\midrule
2000 & 0.2663\,(0.0094) & 0.1625\,(0.0039) & 0.1413\,(0.0034) \\
5000 & 0.2686\,(0.0047) & 0.1654\,(0.0028) & 0.1383\,(0.0024) \\
10000 & 0.2681\,(0.0031) & 0.1668\,(0.0030) & 0.1405\,(0.0031) \\
20000 & 0.2688\,(0.0036) & 0.1637\,(0.0039) & 0.1403\,(0.0036) \\
50000 & 0.2688\,(0.0017) & 0.1635\,(0.0014) & 0.1398\,(0.0024) \\
\bottomrule
\end{tabular}
\end{table}


\bsp	
\label{lastpage}
\end{document}